\documentclass[12pt]{iopart}
\usepackage{epsfig,float,afterpage,amssymb,wrapfig,psfrag}
\newcommand{\be}{\begin{equation}}
\newcommand{\cc}{\mbox{\scriptsize{c}}}
\newcommand{\ii}{\mbox{\scriptsize{i}}} 

\newcommand{\ee}{\end{equation}}
\newcommand{\bea}{\begin{eqnarray}}

\newcommand{\eea}{\end{eqnarray}}
\newcommand{\bra}{\langle}
\newcommand{\spec}{\pi\Delta_0D(\w) \mbox{ vs } \tilde{\w}=\w/\Delta_0}
\newcommand{\ket}{\rangle}
\newcommand{\ssz}{\scriptsize}
\renewcommand{\rmd}{\mbox{d}}
\newcommand{\Atype}{\mbox{\ssz{A}}}
\newcommand{\Btype}{\mbox{\ssz{B}}}
\newcommand{\ei}{\epsilon_{\ii}}
\newcommand{\scrG}{{\cal G}}

\newcommand{\w}{\omega}
\newcommand{\K}{\mbox{\scriptsize{K}}}
\newcommand{\m}{\mbox{\scriptsize{m}}}
\newcommand{\im}{\mbox{i}}
\newcommand{\subi}{\mbox{\scriptsize{I}}}
\newcommand{\subr}{\mbox{\scriptsize{R}}}

\newcommand{\up}{\uparrow}
\newcommand{\eit}{\tilde{\epsilon}_{\ii}}
\newcommand{\eito}{\tilde{\epsilon}_{\ii \mbox{\ssz{o}}}}
\newcommand{\st}{\tilde{\Sigma}}
\newcommand{\ut}{\tilde{U}}
\newcommand{\down}{\downarrow}

\newcommand{\nb}{\bar{n}}
\newcommand{\mb}{\bar{\mu}}
\DeclareMathAlphabet{\bi}{OML}{cmm}{b}{it}
\newcommand{\kk}{\bi{k}}
\newcommand{\sru}{\Sigma^{\subr}_{\up}}
\newcommand{\srd}{\Sigma^{\subr}_{\down}}
\newcommand{\nimp}{n_{\mbox{\ssz{imp}}}}
\newcommand{\srs}{\Sigma^{\subr}_{\sigma}}
\newcommand{\sts}{\tilde{\Sigma}_{\sigma}}

\newcommand{\del}{\Delta_0}

\renewcommand{\theequation}{\arabic{section}.\arabic{equation}}
\newcommand{\ra}{\rightarrow}

\renewcommand{\Or}{\cal{O}\rm}
\newcounter{saveeqn}
\newcommand{\alpheqn}{\setcounter{saveeqn}{\value{equation}}%
\setcounter{equation}{0}%
\addtocounter{saveeqn}{1}%
\renewcommand{\theequation}{\mbox{\arabic{section}.\arabic{saveeqn}\alph{equation}}}%
}
\newcommand{\reseteqn}{\setcounter{equation}{\value{saveeqn}}%
\renewcommand{\theequation}{\arabic{section}.\arabic{equation}}}
\newcommand{\prb}{{\it Phys. Rev. B }}
\newcommand{\jpcm}{{\it J. Phys.: Condens. Matter}}
\newcommand{\prl}{{\it Phys. Rev. Lett. }}
\newcommand{\pr}{{\it Phys. Rev. }}
\newcommand{\rmp}{{\it Rev. Mod. Phys. }}

\newcommand{\seceq}{\setcounter{equation}{0}}

\begin{document}
\jl{31}
\title{Single-particle dynamics of the Anderson model: a local moment approach}
\author{Matthew T Glossop and David E Logan}
\address{Oxford University, Physical and Theoretical Chemistry Laboratory, South Parks Road, Oxford OX1 3QZ, UK}
\tolerance=50
\begin{abstract}
  A non-perturbative local moment approach to single-particle dynamics of the
general asymmetric Anderson impurity model is developed. The approach encompasses
all energy  scales and interaction strengths. It captures thereby strong 
coupling Kondo behaviour, including the resultant universal scaling behaviour of the
single-particle spectrum; as well as the mixed valent and essentially perturbative
empty orbital regimes. The underlying approach is physically transparent and
innately simple, and as such is capable of practical extension to lattice-based models
within the framework of dynamical mean-field theory.
\end{abstract}  

\seceq
\section{Introduction}

The Anderson impurity model [1] (AIM) has long played a pivotal role in
understanding the physical behaviour of materials dominated by strong, local
Coulomb interactions (for a comprehensive review see e.g.\  [2]). A broad range
of problems is encompassed by  the AIM itself, including [2] magnetic impurities
in metals, heavy fermion systems when coherence effects are suppressed, and the 
burgeoning area of quantum dots [3,4]. 

Further impetus to
the study of quantum impurity models has arisen with the advent of dynamical
mean-field theory [5-8] (DMFT), within which correlated lattice-based systems such as
the periodic Anderson or Hubbard models, reduce to an effective quantum impurity 
hybridizing self-consistently with the surrounding fermionic bath. To capture such
problems entails the ability to describe an AIM with essentially arbitrary
dynamics ($\omega$-dependence) in the hybridization function $\Delta(\omega)$ that
embodies coupling between the impurity and the underlying host/bath. The theoretical
difficulties here are considerable, unsurprisingly given the wide range of physics
encompassed; and despite impressive progress in recent years [5-8] there remains a
need for new theoretical approaches that can handle in particular 
the full range of interaction strengths: from non-perturbative strongly correlated 
regimes with their attendant low-energy scales, through to weak coupling, essentially 
perturbative domains of behaviour. 

  Similar comments are in fact applicable even to the conventional metallic AIM [1].
Here, static properties (thermodynamic and related) are certainly well understood
using  a variety of powerful approaches, including the numerical renormalization
group [9], Fermi liquid theory [10] and the Bethe ansatz [11]. But the situation is less
satisfactory when it comes to a theoretical description of dynamics, in particular
single-particle excitations. A wide variety of theories, approximate of necessity,
have of course been
developed in this regard [2]; including for example perturbation theory in the 
interaction strength ($U$) [12], self-consistent renormalization thereof [13], modified
perturbation theory [14], large-$N$ expansions [15,16], the non-crossing approximation
[17-19] and generalizations of it to finite-$U$ [20,21], slave boson mean-field theory
[22,23], a conserving t-matrix approximation [24] and the spinon approximation [25,26]. Their undoubted virtues notwithstanding
however, many of these approaches have well recognized qualitative limitations [2]; 
and there is much scope for further theoretical development, ideally via an approach
that is sufficiently general and practicable that it can be extended with
relative ease to handle correlated lattice-based models within the DMFT framework.

We have recently initiated development of one such, the local moment approach
(LMA) [27-32]. This non-perturbative method is technically simple and 
transparent, with the physically intuitive notion of local moments [1] introduced
explicitly and self-consistently from the outset. For the symmetric AIMs in which
context it has thus far been considered, the LMA handles single-particle
dynamics for all interaction strengths, and on all energy scales including 
recovery of Fermi liquid behaviour at low energies. It captures in particular
the spin-fluctuation physics symptomatic of the strong coupling Kondo regime, manifest
in an exponentially narrow Kondo resonance in the single-particle spectrum $D(\omega)$ 
[27]; such that the resultant scaling behaviour $D(\omega) \equiv F(\omega/\omega_{\K})$
(with $\omega_{\K}$ the Kondo scale) can be obtained in closed form [28], and gives
excellent agreement with NRG calculations [33] that provide benchmark results against
which to compare approximate theories. The LMA has also been extended to finite-$T$
[31], encompassing both single-particle dynamics and associated transport properties 
such as the resistivity. The role of a magnetic field $H$ [30], which poses particular
difficulties for conventional theoretical approaches to dynamics, can likewise be addressed;
and for the Kondo regime in particular, corresponding static properties [29] such as the
Wilson ratio, impurity magnetization and spin susceptibility, are found to agree well
for essentially all field strengths with exact results known e.g.\  from the Bethe
ansatz. Finally but importantly, we add that the LMA is not restricted to the
Fermi liquid physics arising ubiquitously in the metallic AIM. The soft-gap AIM [34],
containing an underlying quantum phase transition between generalized Fermi liquid and
degenerate local moment phases, provides a particular example; in which a rich
range of behaviour has been uncovered by the LMA [32] and confirmed by NRG
calculations [33].

  The LMA considered hitherto is nonetheless specific to particle-hole symmetric
AIMs where the impurity orbital energy, $\epsilon_{\ii}$, is slaved to the on-site
interaction via $\epsilon_{\ii} =-\frac{U}{2}$. It is obviously desirable to extend
the approach to encompass the generic asymmetric case, where the enlarged parameter
space spans a wider range of physical behaviour [2], from the strong coupling Kondo 
domain through mixed valent behaviour to the ultimately perturbative empty orbital
regime. This is important also from the viewpoint of extension to lattice-based models
within DMFT [5-8], where asymmetry in the underlying impurity model corresponds to 
lattice-models away from half-filling; and is hence required to capture e.g.\ 
heavy fermion physics employing the periodic Anderson model.

 A local moment approach to the asymmetric AIM is considered in the present
paper. Following the requisite background (\S2), the two-self-energy description
inherent to the LMA is considered in \S3, with particular emphasis 
on the notion of (self-consistent) symmetry restoration that is central to the 
approach. That discussion is general, being applicable to an essentially 
arbitrary diagrammatic approximation for the underlying self-energies,
and not dependent upon the symmetry-specific arguments hitherto employed for the
symmetric AIM [27,28]. The particular non-perturbative approximation to the LMA 
self-energies implemented here in practice is discussed in \S4.1; it passes the
criterion of practicability, yet appears to capture rather well the relevant regimes
of behaviour. Results arising therefrom are given in \S5, including the rather
subtle issue of universal spectral scaling in the strong coupling Kondo limit (\S5.1);
as well as evolution of single-particle spectra from strong to weak coupling behaviour
(\S5.2) as the Kondo, mixed valence and empty orbital regimes are traversed. The paper
concludes with a brief summary/outlook.

\seceq
\section{Background}
With the Fermi level taken as the energy origin, the familiar Hamiltonian for the spin-$\frac{1}{2}$ AIM [1,2] is 
\be
\fl \ \ \ \ \ \hat{H}=\sum_{\kk,\sigma}\epsilon_{\kk}\hat{n}_{\kk\sigma}+\sum_{\sigma}(\epsilon_{\ii}+\mbox{$\frac{1}{2}$}U\hat{n}_{\ii -\sigma})\hat{n}_{\ii\sigma}+\sum_{\kk, \sigma}V_{\ii\kk}(c^{\dagger}_{\ii\sigma}c_{\kk\sigma}+c^{\dagger}_{\kk\sigma}c_{\ii\sigma}).
\ee
The first term describes the non-interacting host with dispersion $\epsilon_{\kk}$, and the third is the one-electron host-impurity coupling.  The second term refers to the correlated impurity with site-energy $\ei$  and on-site interaction $U$; and the large parameter space of the model is conveniently specified by the asymmetry parameter [1]
\be
\eta=1+\frac{2\ei}{U}
\label{eta}
\ee
with $\eta=0$ for the particle-hole symmetric case (for which $\ei=-\frac{U}{2}$, and the 
impurity charge $n=\sum_{\sigma}\bra\hat{n}_{\ii\sigma}\ket=1$ for all $U$).

Our interest is in single-particle dynamics embodied in the impurity Green function
$G(\w)(\leftrightarrow G(t) =-\im\bra\hat{T}(c_{\ii\sigma}(t)c^{\dagger}_{\ii\sigma})\ket)$, which is naturally independent of spin $\sigma$ ($= \up/\down$ or $+/-$) since $\hat{H}$ is invariant under $\sigma\leftrightarrow -\sigma$; and hence the impurity spectrum $D(\w)=-\frac{1}{\pi}\mbox{sgn}(\w)\mbox{Im}G(\w)$.  In the trivial non-interacting limit $U=0$, the Green function reduces to
\be
g(\w)=\left[\w^+-\ei-\Delta(\w)\right]^{-1}
\label{nig}
\ee
where $\w^+=\w+\im 0^+\mbox{sgn}(\w)$, and is determined solely by the hybridization $\Delta(\w)=\Delta_{\subr}(\w)-\im\mbox{sgn}(\w)\Delta_{\subi}(\w)$ given by 
\be
\Delta(\w)=\sum_{\kk}\frac{\mid V_{\ii\kk}\mid^2}{\w^+-\epsilon_{\kk}}.
\ee
   The essence of the conventional AIM is that the host is metallic  by presumption, corresponding to a non-zero hybridization strength defined by $\del=\Delta_{\subi}(\w=0)$ (with $\w=0$ the Fermi level).  In practice, and without any essential limitation, we thus consider the usual wide flat-band host [2] for which  $\Delta(\w)=-\im\mbox{sgn}(\w)\Delta_0$; where the hybridization strength  is related by $\Delta_0=\pi V^2\rho$ to the host density of states $\rho$ and the matrix elements $V \equiv V_{\ii\kk}$.  The model is thus characterized by two independent parameters: either
\be
\tilde{\epsilon}_{\ii}=\frac{\ei}{\Delta_0} \hspace{2cm} \tilde{U}=\frac{U}{\Delta_0}
\ee
as is traditional; or, as may prove more convenient (and in fact essential in the Kondo scaling regime as we show later), by fixed asymmetry $\eta$ and one or other of $\tilde{U}$, $\tilde{\epsilon}_{\ii}$.

For $U>0$, $G(\w)$ is conventionally written as
\be
G(\w)=\left[g(\w)^{-1}-\Sigma(\w)\right]^{-1}
\label{Gdef}
\ee
in terms of the single self-energy $\Sigma(\w)=\Sigma^{\subr}(\w)-\im\mbox{sgn}(\w)\Sigma^{\subi}(\w)$, with $\Sigma^{\subr}/\Sigma^{\subi}$ related by Hilbert transformation.  The limiting low-$\w$ behaviour of $G(\w)$ that is symptomatic of the Fermi liquid character of the AIM, is the familiar quasiparticle form [2]
\be
G(\w)=\frac{1}{\w^+/Z-\epsilon_{\ii}'+\im\mbox{sgn}(\w)\left[\Delta_0+\Or(\w^2)\right]}
\label{qpform}
\ee
that is simply the leading low-$\w$ expansion of equation (\ref{Gdef}) using $\Sigma^{\subi}(\w)\sim\Or(\w^2)$; where $Z=\left[1-(\partial\Sigma^{\subr}(\w)/\partial \w)_{\w=0}\right]^{-1}$ is the quasiparticle weight, and $\ei'=\ei+\Sigma^{\subr}(0)$ the renormalized level.  The latter is related to the excess charge $n_{\mbox{\ssz{imp}}}$ induced by addition of the impurity, via the Friedel sum rule [35] (which itself follows directly  from the Luttinger integral theorem [36]); specifically [2]
\be
n_{\mbox{\ssz{imp}}}=2\int_{-\infty}^{0}\rmd\w \ \Delta\rho(\w)=1-\frac{2}{\pi}\mbox{tan}^{-1}\left(\frac{\ei+\Sigma^{\subr}(0)}{\Delta_0}\right)
\label{friedel}
\ee
such that ($\ei'=$) $\ei+\Sigma^{\subr}(0)=\del\mbox{tan}[\frac{\pi}{2}(1-n_{\mbox{\ssz{imp}}})]$.  Here $\Delta\rho(\w)=\rho(\w)-\rho_{\mbox{\ssz{host}}}(\w)$ is the change in total density of states of the system due to addition of the impurity, given by $\Delta\rho(\w)=-\frac{1}{\pi}\mbox{sgn}(\w)\mbox{Im}\left\{G(\w)\left[1-\partial \Delta(\w)/\partial \w\right]\right\}$; so that $n_{\mbox{\ssz{imp}}}$ is related in general to the local impurity charge
\alpheqn
\be
n=2\int_{-\infty}^0\rmd\w\ D(\w)
\ee
by
\be
n_{\mbox{\ssz{imp}}}=n-\frac{2}{\pi}\mbox{Im}\int_{-\infty}^0 \rmd \w\ G(\w)\frac{\partial\Delta(\w)}{\partial \w}
\label{nimpeqn}
\ee
\reseteqn
(with $n_{\mbox{\ssz{imp}}}=n$ for the wide flat-band case considered in practice).  Equations (\ref{qpform},8) lead directly in turn to 
\be
\pi\Delta_0D(\w=0)=\mbox{sin}^2\left(\mbox{$\frac{\pi}{2}$}n_{\mbox{\ssz{imp}}}\right)
\ee
relating the Fermi level spectrum to $n_{\mbox{\ssz{imp}}}$, and reducing to the `pinning' condition $\pi\del D(0)=1$ for the particle-hole symmetric case where $n_{\mbox{\ssz{imp}}}=1$ always.  The above results, all well known, will prove important in subsequent sections.

The quasiparticle form equation (\ref{qpform}), conjoined with the Friedel sum rule for $\ei'=\ei+\Sigma^{\subr}(0)$, forms a starting point for microscopic Fermi liquid theory as well as being the essential result for $G(\w)$ arising from slave boson mean-field theory (see e.g.\ [2]).  Its familiarity should not however obscure its limitations, for it is confined to the lowest frequencies $\w/\Delta_0Z\ll 1$; and as such (save trivially for $U=0$) captures only a tiny fraction [28] of the Abrikosov-Suhl resonance, let alone high-energy spectral features such as the Hubbard satellites.  To capture dynamics on all energy scales, while recovering correctly the limiting quasiparticle form equation (\ref{qpform}), is not a trivial matter; and non-perturbative approaches are certainly required to handle e.g.\ the strong coupling Kondo regime.  One such is discussed here.

Before proceeding, two remaining relevant points should be made.  First, briefly, to specify the minimal range of asymmetry $\eta$ (equation (\ref{eta})) that need be considered.  Under a particle-hole transformation, $\ei$ is replaced by $-[\ei+U]$ [9]; and, labelling temporarily the $\ei$-dependence of $G(\w)$ and $n_{\mbox{\ssz{imp}}}$, it is straightforward to show that $G(\w; \ei)=-G(-\w; -[\ei+U])$, i.e.\ 
\be
D(\w;\ei)=D(-\w;-[\ei+U])
\label{eisym}
\ee
and hence $n_{\mbox{\ssz{imp}}}(\ei)=n_{\mbox{\ssz{imp}}}(-[\ei+U])$.  From this it follows that only $\ei\ge -U/2$ (and hence $0\le n_{\mbox{\ssz{imp}}} \le 1$) need be considered, corresponding (equation (\ref{eta})) to $\eta\ge 0$.  This range is assumed from now on.

The second point refers to the strong coupling Kondo limit of the AIM.  Corresponding to $n_{\mbox{\ssz{imp}}}\ra 1$, this arises when $\ei < 0$ such that $|\eit|\gg 1$ and $\ut-|\eit|\gg 1$ (i.e.\ the lower/upper Hubbard satellites centred on $\ei=-|\ei|$ and $\ei+U$ respectively are well below/above the Fermi level).  The approach to the Kondo limit is not therefore unique, in that $n_{\mbox{\ssz{imp}}}\ra 1$ arises for {\it any} given asymmetry $\eta\equiv 1-2|\eit|/\ut \in [0,1]$  upon progressively increasing either $\ut$ or $|\eit|$.  This is reflected in turn by the fact that the Kondo model onto which the low-energy sector of the AIM maps in strong coupling under a Schrieffer-Wolff transformation [37], contains both exchange ($J$) {\it and} potential ($K$) scattering contributions [2]; viz
\be
\hat{H}_{\mbox{\ssz{\K}}}=\sum_{\kk,\sigma}\epsilon_{\kk}\hat{n}_{\kk\sigma}+2J\hat{{\bf s}}_{\ii}\cdot\hat{{\bf S}}(0)+\frac{1}{2}K\sum_{\kk,\kk',\sigma}c^{\dagger}_{\kk\sigma}c_{\kk'\sigma}.
\label{kondo}
\ee
Here $\hat{\bf{s}}_{\ii}$ and $\hat{{\bf S}}(0)$ denote respectively the impurity spin and conduction electron spin density at the impurity; and the exchange and potential scattering matrix elements are related to the bare parameters of the AIM by [2]:
\alpheqn
\bea
\rho J&=&\frac{\del}{\pi}\left\{\frac{1}{|\ei|}+\frac{1}{U-|\ei|}\right\}=\left[\frac{\pi|\ei|(U-|\ei|)}{\del U}\right]^{-1} \label{rhoj}\\
\rho K &=&\frac{\del}{\pi}\left\{\frac{1}{|\ei|}-\frac{1}{U-|\ei|}\right\}.
\eea
\reseteqn
The appropriate Kondo model may thus be characterized by the two dimensionless parameters $\rho J$ and $K/J$, with $\rho J \ll 1$ for the AIM $\ra$ Kondo reduction to be valid; and where from equations (\ref{rhoj},b) and (\ref{eta})
\be
\frac{K}{J}=1-\frac{2|\ei|}{U}\equiv \eta
\ee
is simply the asymmetry of the underlying AIM.  Note that the Kondo model that in practice is usually considered lacks potential scattering, and hence corresponds to the strong coupling limit of the symmetric AIM alone ($\eta=0$).  In general however, physical properties of the AIM in the strong coupling regime should depend on $K/J=\eta$; a statement that, with one limiting exception, applies in particular to the scaling behaviour of the single-particle spectrum, $D(\w)\equiv F(\w/\w_{\K})$ with $\w_{\K}\propto \Delta_0 Z \ra 0$ the Kondo scale.  That exception resides in the {\it leading} low-$\w$ behaviour of $G(\w)$ embodied in the quasiparticle form $G(\w)$, equation (\ref{qpform}); which in the Kondo limit where $n_{\mbox{\ssz{imp}}}\ra 1$ (and hence $\ei'=\ei+\Sigma^{\subr}(0)\ra 0$), is dependent solely upon $\w/\Delta_0Z \propto \w/\w_{\K}$ with no explicit dependence on $K/J=\eta$.  To our knowledge one nonetheless has no {\it a priori} reason to expect that the single-particle scaling spectrum on {\it all} energy scales $\w/\w_{\K}$ will be independent of the asymmetry $\eta$.

\seceq
\section{LMA: basis}
The conventional route to single-particle dynamics is via the usual `single' self-energy $\Sigma(\w)$.  But this is merely defined by the Dyson equation implicit in equation (\ref{Gdef}), and a determination of $G(\w)$ in this way is not obligatory; indeed there may be good reasons to avoid it, notably the inability of conventional perturbation theory to handle strong correlations in general.  The LMA thus eschews such an approach, focusing instead on a two-self-energy description that is a natural consequence of the mean-field approach from which it starts.  In this section we consider briefly the implications of such, in a manner that is neither dependent on the essential details of practical implementation (\S 4.1) nor confined to the particle-hole symmetric case considered hitherto [27-32].

There are three essential elements to the LMA.  (i) First that local moments (`$\mu$'), viewed as the primary effect of interactions, are introduced explicitly from the outset.  As in Anderson's original work [1], the {\it starting} point is thus simple static mean-field (MF), i.e.\ unrestricted Hartree Fock.  This contains in general two {\it degenerate} broken symmetry MF states (reflecting the invariance of $\hat{H}$ under $\sigma \leftrightarrow -\sigma$); denoted by $\alpha=$ A or B and corresponding respectively to local moments $\mu=+|\mu|$ and $-|\mu|$.  Notwithstanding the severe limitations of MF by itself, it may nonetheless be used as a basis for a genuine many-body approach encompassing the correlated electron dynamics that are the essence e.g.\ of Kondo physics.  (ii) To this end the LMA employs the two-self-energy description that follows naturally from the underlying {\it two} mean-field saddle points; with non-trivial dynamics introduced in practice (\S 4.1) into the associated self-energies via coupling of single-particle excitations to low-energy transverse spin fluctuations.  (iii) The final, central notion behind the LMA is that of symmetry restoration: self-consistent restoration of the broken symmetry endemic at mean-field level, and recovery of Fermi liquid behaviour, as pursued below.

Within the LMA, $G(\w)$ is expressed formally as 
\alpheqn
\be
G(\w)=\frac{1}{2}\sum_{\alpha}G_{\alpha\sigma}(\w)
\label{sumalpha}
\ee
with propagators $G_{\alpha\sigma}(\w)=[g^{-1}(\w)-\tilde{\Sigma}_{\alpha\sigma}(\w)]^{-1}$ built from the appropriate MF state $\alpha=$ A or B, and self-energies separated as $\tilde{\Sigma}_{\alpha\sigma}(\w)=\tilde{\Sigma}^0_{\alpha\sigma}+\Sigma_{\alpha\sigma}(\w)$ into a purely static contribution $\tilde{\Sigma}^0_{\alpha\sigma}$ that alone is retained at pure MF level; together with $\Sigma_{\alpha\sigma}(\w)=\Sigma_{\alpha\sigma}[\{\scrG_{\alpha\sigma}\}]$ that in particular contains the key dynamics, and is a functional of (and built diagrammatically from) the underlying MF propagators $\scrG_{\alpha\sigma}(\w)$.  The first, brief issue here is rotational invariance: the fact that $G(\w)$ is independent of spin, $\sigma$.  This is correctly preserved, since the invariance of $\hat{H}$ under $\sigma\leftrightarrow -\sigma$ implies $G_{\Atype\sigma}(\w)=G_{\Btype-\sigma}(\w)$; whence the sum in equation (\ref{sumalpha}) is indeed $\sigma$-independent.  By the same token, equation (\ref{sumalpha}) may be written as
\be
G(\w)=\frac{1}{2}\sum_{\sigma}G_{\alpha\sigma}(\w)
\label{sumsigma}
\ee
\reseteqn
involving a `spin sum' that is independent of $\alpha$.  Equations (\ref{sumalpha},b) are entirely equivalent.  We choose to work with the latter form and, since the $\alpha$ label is redundant, we drop it from now on.

The impurity Green function is thus 
\alpheqn
\be
G(\w)=\frac{1}{2}\left[G_{\up}(\w)+G_{\down}(\w)\right]
\label{Gupdn}
\ee
with
\be
G_{\sigma}(\w)=\left[g^{-1}(\w)-\tilde{\Sigma}_{\sigma}(\w)\right]^{-1}
\label{Gsig}
\ee
\reseteqn
and self-energies $\tilde{\Sigma}_{\sigma}(\w)=\tilde{\Sigma}^{\subr}_{\sigma}(\w)-\im\mbox{sgn}(\w)\tilde{\Sigma}^{\subi}_{\sigma}(\w)$ separated as 
\be
 \psfrag{xxxxx}[bc][bc]{$\st_{\sigma}(\w)=\ \ $} \psfrag{ooooo}[bc][bc]{$+\ \Sigma_{\sigma}(\w)$} \epsfig{file =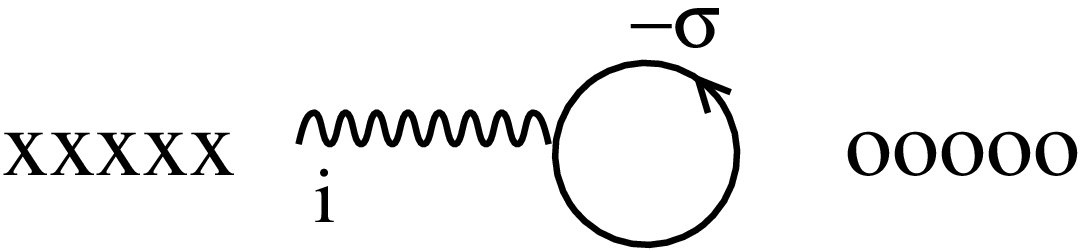,width=4.5cm}
\label{sesep}
\ee
with $\tilde{\Sigma}_{\sigma}^0$ given by the static bubble diagram (and `everything else' in $\Sigma_{\sigma}(\w)$).

The conventional single self-energy $\Sigma(\w)$ follows directly from an underlying two-self-energy description, being given on comparison of equations (\ref{Gdef}) and (\ref{Gupdn},b) by
\be
\Sigma(\w)=\mbox{$\frac{1}{2}$}\left[\tilde{\Sigma}_{\up}(\w)+\tilde{\Sigma}_{\down}(\w)\right]+\frac{\left[\frac{1}{2}(\tilde{\Sigma}_{\up}(\w)-\tilde{\Sigma}_{\down}(\w))\right]^2}{g^{-1}(\w)-\frac{1}{2}(\tilde{\Sigma}_{\up}(\w)+\tilde{\Sigma}_{\down}(\w))}
\label{2to1}
\ee
with $g(\w)$ the non-interacting propagator equation (\ref{nig}).  At the pure MF level of unrestricted Hartree Fock (HF), the dynamics of $\tilde{\Sigma}_{\sigma}(\w)$ are neglected and $\tilde{\Sigma}_{\sigma}(\w)\equiv\tilde{\Sigma}_{\sigma}^0=\frac{U}{2}(n-\sigma|\mu|)$ (with the local charge ($n$) and moment ($|\mu|$) determined self-consistently in the usual way, \S 3.1); from equation (\ref{2to1}) the corresponding self-energy is then:
\be
\Sigma_{\mbox{\ssz{HF}}}(\w)=\frac{U}{2}n+\frac{(\frac{1}{2}U|\mu|)^2}{\w^+-\left[\ei+\frac{U}{2}n\right]-\Delta(\w)}.
\label{sigmahf}
\ee
The problem with MF by itself is clear, for if $|\mu|\neq 0$ then $\Sigma^{\subi}(\w=0)\neq 0$ and Fermi liquid behaviour is violated.  This is not surprising, for the resultant degenerate local moment state is not perturbatively connected to the non-interacting limit; but it is no less wrong for that and, despite the physical appeal of local moment formation, is a major reason why MF has not hitherto provided a successful basis for a many-body approach.  This problem does not of course occur if $|\mu|=0$ is enforced (restricted HF), but there another one arises.  The two- and single- self energy descriptions then coincide, and $\Sigma_{\mbox{\ssz{HF}}}(\w)=\frac{U}{2}n (\equiv\tilde{\Sigma}^0_{\sigma})$ is just the static Hartree contribution.  This produces a trivial energy shift to the non-interacting propagator, and subsequent construction of the dynamical $\Sigma(\w)$ via conventional perturbation theory in $U$ is essentially equivalent to expanding about the restricted HF state.  But when local moments can form at MF level, this saddle point --- in contrast to those of unrestricted HF --- is {\it unstable} (a local maximum, with the single restricted HF determinant unstable to particle-hole excitations).  And it is this that underlies at heart the divergences that plague conventional perturbation theory [2] if one attempts to perform the more or less standard diagrammatic resummations (e.g.\ RPA) that one might expect are required to capture the regime of strong correlations. We do not of course doubt the applicability in {\it principle} of perturbation theory in $U$ --- for the metallic AIM; but practice is another matter.

The LMA in a sense seeks the best of both worlds: to retain the two-self-energy description with the notion of local moments (and essential stability of the underlying MF state), while incorporating dynamics into the associated self-energies $\tilde{\Sigma}_{\sigma}(\w)$ in such a way that the resultant description is simple and tractable, and yet recovers Fermi liquid behaviour at low energies.

\subsection{Symmetry restoration}
The first question here is the latter point: under what conditions on the   $\tilde{\Sigma}_{\sigma}(\w)$'s will the single $\Sigma(\w)$ exhibit Fermi liquid behaviour at low-$\w$, i.e.\ $\Sigma^{\subi}(\w)\sim\Or(\w^2)$?  To answer this, consider a simple low-$\w$ expansion of the $\tilde{\Sigma}_{\sigma}(\w)$'s (equation (\ref{sesep})), viz
\alpheqn
\be
 \tilde{\Sigma}^{\subr}_{\sigma}(\w) \sim\ \tilde{\Sigma}^{\subr}_{\sigma}(0)-\left[\frac{1}{Z_{\sigma}}-1\right]\w
\label{realexp}
\ee
for the real parts, where $Z_{\sigma}=[1-(\partial \Sigma^{\subr}_{\sigma}(\w)/\partial\w)_{\w=0}]^{-1}$ is thus defined and no {\it a priori} constraints are imposed on $\tilde{\Sigma}^{\subr}_{\sigma}(0)=\st_{\sigma}^{0}+\Sigma_{\sigma}(\w=0)$ at the Fermi level $\w=0$; together with
\be
(\st^{\subi}_{\sigma}(\w)\equiv\ )\ \Sigma^{\subi}_{\sigma}(\w)\sim a_{\sigma}\w^2
\label{imexp}
\ee
\reseteqn
for the imaginary part.  The latter form is guaranteed from the diagrams (see \S 4) for $\Sigma_{\sigma}(\w)\equiv \Sigma_{\sigma}[\{\scrG_{\sigma}\}]$ provided the host is metallic ($\Delta_0\neq 0$, as appropriate to the metallic AIM), but we emphasize is {\it not} sufficient by itself to guarantee $\Sigma^{\subi}(\w)\sim \Or(\w^2)$.  Equations (\ref{realexp},b), together (if appropriate) with a low-$\w$ expansion of the hybridization $\Delta(\w)$, may then be used in equation (\ref{2to1}) to determine the low-$\w$ behaviour of the resultant single self-energy.  This is a matter of algebra, and we find thereby that the necessary/sufficient condition for 
\be
\Sigma^{\subi}(\w)\sim\Or(\w^2)
\ee
is that:
\be
\tilde{\Sigma}_{\up}^{\subr}(\w=0)=\tilde{\Sigma}_{\down}^{\subr}(\w=0)
\label{sr}.
\ee
If equation (\ref{sr}) is satisfied, then from equations (\ref{2to1}) and (\ref{realexp},b) (i) all self-energies coincide at the Fermi level,
\be
\Sigma^{\subr}(\w=0)=\st^{\subr}_{\sigma}(\w=0)
\ee
(for either $\sigma$); (ii) the low-$\w$ behaviour of $\Sigma^{\subr}(\w)$ is
\be
\Sigma^{\subr}(\w)\sim\Sigma^{\subr}(0)-\left(\frac{1}{Z}-1\right)\w
\ee
with the usual quasiparticle weight $Z=\left[1-(\partial\Sigma^{\subr}(\w)/\partial\w)_{\w=0}\right]^{-1}$ related to the $\{Z_{\sigma}\}$ by $Z^{-1}=\frac{1}{2}(Z^{-1}_{\up}+Z^{-1}_{\down})$; and (iii) the quasiparticle form equation (\ref{qpform}) for $G(\w)$ is recovered.

Equation (\ref{sr}), referring solely to the Fermi level $\w=0$, is the symmetry restoration (SR) condition that is central to the LMA; and which, if satisfied, guarantees Fermi liquid behaviour.  It is quite general, meaning not confined to the particle-hole symmetric AIM considered hitherto  (which is recovered as a particular case of the above, but was originally argued for [27,28] on different, symmetry-specific grounds).   The general consequences of SR in practice are nonetheless as found for the symmetric model [27,28] : Self-consistent imposition of equation (\ref{sr}) amounts (\S 4) to a self-consistency condition for the local moment $|\mu|$ and, most importantly, generates a low-energy spin-flip scale $\w_{\m}$.  The latter, manifest in particular as a strong resonance in the transverse spin polarization propagator $\mbox{Im}\Pi^{+-}(\w)$, is the Kondo scale, exponentially small in strong coupling; its physical significance in the approach being that it sets the timescale $\tau\sim h/\w_{\m}$ for restoration of the broken symmetry inherent at crude MF level (and arising in effect from dynamical tunneling between the degenerate MF minima).  We also add that if the SR condition equation (\ref{sr}) cannot be satisfied, then a doubly degenerate local moment phase  results [32] (with a characteristic spin-flip scale $\w_{\m}=0$ reflecting the local degeneracy, and hence $\tau=\infty$).  While this does not arise for the metallic AIM where SR is ubiquitously satisfied, it is the self-consistent possibility of such embodied in SR that enables the LMA to capture [32] e.g.\ the quantum phase transition from a (generalized) Fermi liquid to a local moment state in the soft-gap AIM ($\Delta_{\subi}(\w)\sim|\w|^r$) [34] where both phases arise.

Finally, note that with the SR condition equation (\ref{sr}) satisfied, the Friedel sum rule equation (\ref{friedel}) may be written as
\be
\ei+\st^{\subr}_{\sigma}(\w=0)=\del\mbox{tan}\left[\frac{\pi}{2}(1-n_{\mbox{\ssz{imp}}})\right]
\label{fsr}
\ee
(independent of $\sigma$).   And for later use, the LMA $G_{\sigma}(\w)$'s may then be expressed using equations (\ref{Gsig}) and (\ref{sesep}) as
\alpheqn
\be
\fl \ \ \ G_{\sigma}(\w)=\left[\w^+-\Delta(\w)-\Delta_0\mbox{tan}\left[\frac{\pi}{2}(1-n_{\mbox{\ssz{imp}}})\right]-\left(\Sigma_{\sigma}(\w)-\Sigma_{\sigma}(0)\right)\right]^{-1}
\label{Gofn}
\ee 
being dependent only on the `non-MF' contributions ($\Sigma_{\sigma}(\w)$) to $\st_{\sigma}(\w)$; such that $n_{\mbox{\ssz{imp}}}$ is given via equation (2.9) (using equation (3.2)) by
\be
n_{\mbox{\ssz{imp}}}=\sum_{\sigma}\mbox{Im}\int_{-\infty}^0\frac{\rmd \w}{\pi} \ G_{\sigma}(\w)\left[1-\frac{\partial\Delta(\w)}{\partial \w}\right].
\ee
\reseteqn 
The Friedel sum rule is not however satisfied by approximate theories in general, save for the particle-hole symmetric case where it is guaranteed by symmetry.  That is true e.g.\ of the NCA [38], and even for conventional second order perturbation theory in $U$ [12], where $n_{\mbox{\ssz{imp}}}$ inferred from $\ei+\Sigma^{\subr}(0)=\del\tan [\frac{\pi}{2}(1-n_{\mbox{\ssz{imp}}})]$ (equation (\ref{friedel})) using the second-order $\Sigma^{\subr}(0)$ does not coincide with $n_{\mbox{\ssz{imp}}}$ obtained from spectral integration.  Satisfaction of the Friedel sum rule, which is distinct from the issue of symmetry restoration, is however desirable if possible [14]; and in \S 4 we also show how to incorporate it naturally within the LMA.  The Friedel sum rule is also satisfied in the modified perturbation scheme [14], where the single self-energy (excluding the Hartree piece) is parametrized in the form $\Sigma(\w)\simeq A\Sigma_0^{(2)}(\w)[1-B\Sigma_0^{(2)}(\w)]^{-1}$; with $\Sigma_0^{(2)}(\w)$ the second order self-energy constructed from non-interacting Green functions containing a shifted chemical potential $\tilde{\mu}_0$.  The parameters $A,B,\tilde{\mu}_0$ are then found [14] by requiring that $\Sigma(\w)$ has the correct high-frequency behaviour (in $1/\w$), that the atomic limit is recovered, and that the Friedel sum rule is satisfied.  While simple and practicable, the approach is however too crude to handle e.g.\ the Kondo regime (as evident for example in the fact that for the symmetric AIM where $B=0$ and $A=1$, $\Sigma(\w)\simeq \Sigma_0^{(2)}(\w)$ reduces to straight second order perturbation theory and hence produces a Kondo scale that is algebraically rather than exponentially small in $\ut$). 

\subsection{Mean field}
Before proceeding we reprise briefly elements of pure MF that will be required in the following sections (full details are given in [1]).  The MF propagators $\scrG_{\sigma}(\w)$ may be expressed as $\scrG_{\sigma}(\w)=[\w^+-\epsilon_{\ii\sigma}-\Delta(\w)]^{-1}$ with $\epsilon_{\ii\sigma}=\ei+\st^0_{\sigma}=\ei+\frac{U}{2}(n-\sigma|\mu|)$; and $n,|\mu|$ found self-consistently at pure MF level via $\bra \hat{n}_{\ii\up}\pm \hat{n}_{\ii\down}\ket_0$ respectively (with $\bra...\ket_0$ a MF average).  The corresponding spectra $D^0_{\sigma}(\w)\equiv D^0_{\sigma}(\w;e_{\ii},x)$ are given (for the wide-band host explicitly) by
\be
D^0_{\sigma}(\w)=\frac{\Delta_0\pi^{-1}}{[\w-e_{\ii}+\sigma x]^2+\Delta_0^2}
\ee
where $x=\frac{1}{2}U|\mu|$ and $e_{\ii}=\ei+\frac{U}{2}n$.  The MF charge and moment are thus found from self-consistent solution of
\alpheqn
\be
|\mu|=\sum_{\sigma}\sigma\int_{-\infty}^0 \rmd \w \ D^0_{\sigma}(\w;e_{\ii},x)
\label{UHFmoment}
\ee
\be
n=\sum_{\sigma}\int_{-\infty}^0 \rmd \w \ D^0_{\sigma}(\w;e_{\ii},x).
\label{UHFcharge}
\ee
\reseteqn
We choose for later convenience to work at fixed $U$; in which case the MF $|\mu|$ follows by solution of equation (\ref{UHFmoment}) for given $e_{\ii}$, the charge $n$ follows directly from equation (\ref{UHFcharge}), and the corresponding `bare' $\ei$ then follows immediately from $\ei=e_{\ii}-\frac{U}{2}n$.

The boundary to local moment formation at MF level is readily deduced from equations (\ref{UHFmoment},b) and is given in closed from by
\be
\eta_{\cc}=\pm\frac{2}{\pi}\left[\mbox{tan}^{-1}\left(\sqrt{\tilde{U}'_{\cc}-1}\ \right)+\frac{\sqrt{\tilde{U}'_{\cc}-1}}{\tilde{U}'_{\cc}}\right]
\ee
(which recovers figure 4 of [1]), where $\tilde{U}'=\tilde{U}/\pi$ and $\eta$ is the asymmetry (equation (\ref{eta})); such that the MF $|\mu|>0$ for any given $|\eta|<1$ when $\tilde{U}'>\tilde{U}'_{\cc}$ (or equivalently for all $|\eta|<|\eta_{\cc}|$ and any given $\tilde{U}'>1$).

\seceq

\section{LMA: practice}
\label{lmase}

In this section we consider first the approach to solving the problem  generally within the LMA framework, for an essentially arbitrary approximation to the self-energies $\tilde{\Sigma}_{\sigma}(\w)$; before turning in \S 4.1 to the specific class of diagrams retained in practice, and the key consequences of stability and symmetry restoration.

The LMA self-energies $\st_{\sigma}(\w)$ (equation (3.3)) are functionals of the MF propagators, viz $\scrG_{\sigma}(\w)=[\w^+-e_{\ii}+\sigma x -\Delta(\w)]^{-1}$ (with corresponding spectral density $D^0_{\sigma}(\w)\equiv D^0_{\sigma}(\w;e_{\ii},x)$, equation (3.13)).  With interactions included beyond MF level, $x=\frac{1}{2}U|\mu|$ and $e_{\ii}$ are determined self-consistently as below and naturally differ from their pure MF values; so for clarity the latter are denoted from now on as $x_0 =\frac{1}{2}U|\mu_0|$ and $e_{\ii}^0=\ei+\frac{U}{2}n_0$ (as obtained from the pure MF equations (3.14)).  Equation (3.3) for $\st_{\sigma}(\w)$ is given by
\be
\st_{\sigma}(\w)=\frac{U}{2}[\bar{n}-\sigma|\bar{\mu}|]+\Sigma_{\sigma}(\w)
\label{sedecomp}
\ee
where the first term is the static bubble diagram, $\st^0_{\sigma}=U\int_{-\infty}^0\rmd\w \ D^0_{-\sigma}(\w; e_{\ii},x)$; and $|\mb|\equiv |\mb(e_{\ii},x)|$ and $\nb\equiv\nb(e_{\ii},x)$ are given by
\alpheqn
\be
|\mb|=\sum_{\sigma}\sigma\int_{-\infty}^0\rmd\w\ D_{\sigma}^0(\w;e_{\ii},x)
\label{mubar}
\ee
\be
|\nb|=\sum_{\sigma}\int_{-\infty}^0\rmd\w\ D_{\sigma}^0(\w;e_{\ii},x).
\label{nbar}
\ee
\reseteqn
It is of course the `post-MF' contribution, $\Sigma_{\sigma}(\w)$, that is all important; and a suitable, naturally approximate choice for which determines the extent to which the key physics of the problem is captured in practice.  That is considered in \S 4, but for the present assume $\Sigma_{\sigma}(\w)\equiv\Sigma_{\sigma}[\{\scrG_{\sigma}(\w)\}]$ to be given.  It depends generally on $U$ (via the diagrammatic interaction vertices) and, since it is a functional of $\{\scrG_{\sigma}\}$, upon $e_{\ii}$ and $x$; i.e.\ $\Sigma_{\sigma}(\w)\equiv\Sigma_{\sigma}(\w; e_{\ii},x)$ (with the U-dependence implicit).  The symmetry restoration condition equation (\ref{sr}) is then of form:
\be
\sru(\w=0;e_{\ii},x)-\srd(\w=0;e_{\ii},x)=U|\mb(e_{\ii},x)|.
\label{srnew}
\ee
And the Friedel sum rule equation (\ref{fsr}) (equivalent to the Luttinger theorem [36]) is given by
\be
\fl \ \ \ \ \  \ei+\frac{U}{2}[\nb(e_{\ii},x)-\sigma|\mb(e_{\ii},x)|]+\srs(\w=0;e_{\ii},x)=\Delta_0\tan [\mbox{$\frac{\pi}{2}$}(1-n_{\mbox{\ssz{imp}}})]
\label{fsrnew}
\ee
(independently of $\sigma$).

The problem may then be solved in the following natural way, considered for (any) fixed $U$:
\begin{enumerate}
\itemsep -0.3cm
\item For given $e_{\ii}$ [or $x]$, the symmetry restoration condition equation (\ref{srnew}) is solved for $x=\frac{1}{2}U|\mu|$ [or $e_{\ii}$].\\
\item $n_{\mbox{\ssz{imp}}}$ is then obtained by solution of equations (3.12).\\
\item All quantities in equation (\ref{fsrnew}) save $\ei$ are then known, whence the bare $\ei$ follows directly from equation (\ref{fsrnew}).
\end{enumerate}
Alternatively, if one wishes to specify a bare $\ei$ from the outset, the procedure may be repeated by varying $e_{\ii}$ until equation (\ref{fsrnew}) is satisfied.  The method above is straightforward, and in particular we remind the reader that (see \S 3) satisfaction of symmetry restoration (step (i)) guarantees the requisite low-energy Fermi liquid behaviour.  As considered in practice (\S 4.1ff) it is also numerically rapid, with each of the iterative steps (i) and (ii) typically found to converge after a small number of iterations.

Using the  symmetry $\scrG_{\sigma}(\w;e_{\ii},x)=-\scrG_{-\sigma}(-\w;-e_{\ii},x)$ together with the general diagrammatic structure of  $\st_{\sigma}(\w)\equiv\st_{\sigma}[\{\scrG_{\sigma}\}]$, it can moreover be shown that $G_{\sigma}(\w;e_{\ii},x)=-G_{-\sigma}(-\w;-e_{\ii},x)$ and $\st_{\sigma}(\w;e_{\ii},x)=U-\st_{-\sigma}(-\w;-e_{\ii},x)$; and in consequence that if $(x,n_{\mbox{\ssz{imp}}},\ei)$ is a solution of the above equations for $\e_{\ii}=\alpha$, then $(x,2-n_{\mbox{\ssz{imp}}},-[\ei+U])$ is also a solution for $e_{\ii}=-\alpha$.  In other words the particle-hole transformation mentioned in \S 2, under which $\ei\ra -[\ei+U]$  and $\nimp \ra 2-\nimp$, corresponds to $e_{\ii}\ra -e_{\ii}$ (with $D_{\sigma}(\w;\ei) = D_{-\sigma}(-\w;-[\ei+U])$ such that equation (2.11) follows directly, since $D(\w)=\frac{1}{2}\sum_{\sigma}D_{\sigma}(\w)$).  From this it follows in turn that only $e_{\ii}\ge 0$ need be considered (corresponding to $\nimp\le 1$); and that the particle-hole symmetric AIM, for which $\ei=-\frac{U}{2}$ and $\nimp=1$, corresponds to $e_{\ii}=0$.  For the latter case, steps (ii) and (iii) above are redundant, the Friedel sum rule equation (\ref{fsrnew}) being guaranteed by particle-hole symmetry since $\st^{\subr}_{\sigma}(\w=0)=\frac{U}{2}$ in that case; and solely the central symmetry restoration condition equation (\ref{srnew}) need be solved.  The problem then becomes that considered previously [27,28] (noting that $\st_{\sigma}(\w)$ used therein is defined to exclude the Hartree contribution of $\frac{U}{2}$), with one minor exception; namely that in [27-32] $|\mu|$ itself rather than $|\mb|$ was employed on the right hand side of equation (\ref{srnew}).  (The latter, used here, is technically correct; although we have confirmed that the results of [27-32] are entirely unaffected by this replacement.)

\subsection{Self-energies}
The specific class of diagrams contributing to  $\Sigma_{\sigma}(\w)\equiv\Sigma_{\sigma}[\{\scrG_{\sigma}\}]$ which we retain in practice is precisely that considered hitherto for the symmetric AIM [27-32]; shown in figure 1 and translating to 
\be
\Sigma_{\sigma}(\w)=U^2\int_{-\infty}^{\infty}\frac{\rmd\w_1}{2\pi\im}\ \scrG_{-\sigma}(\w-\w_1)\Pi^{-\sigma\sigma}(\w_1).
\label{se}
\ee
Physically, these diagrams capture dynamical spin-flip scattering processes: in which having, say, added a $\sigma$-spin electron to a $-\sigma$-spin occupied impurity, the latter hops off the impurity  and thus generates an on-site spin-flip (reflected in the transverse spin polarization propagator $\Pi^{-\sigma\sigma}(\w)$), before returning again at a later time.  At the simplest level, likewise considered here, the polarization propagator is given by an RPA-like particle-hole ladder sum in the transverse spin channnel; specifically
\be
\Pi^{\sigma-\sigma}(\w)=\ ^0\Pi^{\sigma-\sigma}(\w)[1-U ^0\Pi^{\sigma-\sigma}(\w)]^{-1}
\label{fullpi}
\ee
with $^0\Pi^{\sigma-\sigma}(\w)$ the bare particle-hole bubble, itself expressed in terms of the broken symmetry MF propagators $\{\scrG_{\sigma}\}$.  Further details regarding the $\Sigma_{\sigma}(\w)$ diagrams and their physical content are given in [27,32] (see e.g.\ figure 3 of [32]).  The full LMA self-energies $\tilde{\Sigma}_{\sigma}(\w)$ follow directly from equations (\ref{sedecomp},5).

\begin{wrapfigure}{r}{5cm}
\epsfig{file =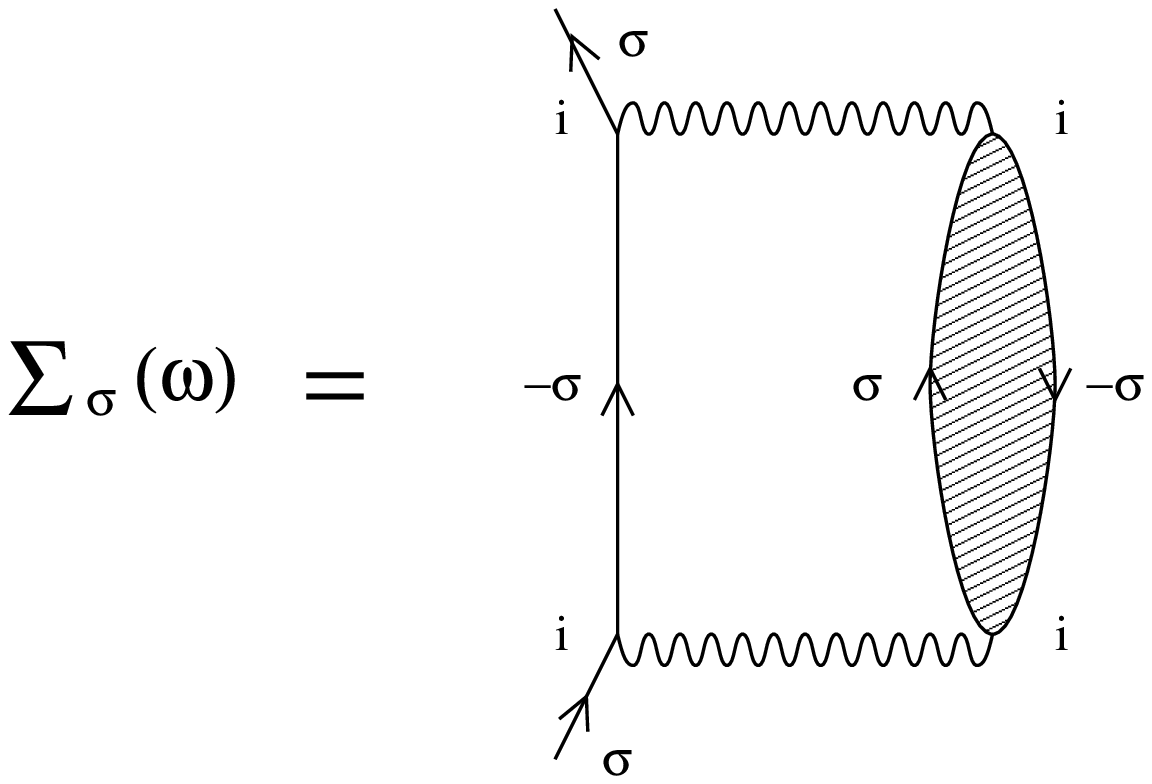,width=5cm} 
\label{static}
\caption{}
{\small Principal contribution to the LMA $\Sigma_{\sigma}(\w)$, see text.  Wavy lines denote U.}
\end{wrapfigure}

Retention of the above diagrams for $\Sigma_{\sigma}(\w)$ is motivated primarily on physical grounds, since they embody the dynamical coupling of single-particle excitations to low-energy transverse spin fluctuations that is essential to capture in particular the strong coupling Kondo regime.  Other classes of diagrams are readily retained, notably those shown in figure 9 of [27] involving repeated particle-particle interactions in the transverse spin channel, as well as the repeated `bubble' sum.  But we find these to have little effect on the results given here, and that the class retained appears to handle rather well essentially all regimes of the AIM, from Kondo to empty orbital; reflecting at least in part the fact that the present approach asymptotically recovers straight second order perturbation theory in $U$ in the weak coupling domain appropriate to the empty orbital regime (as will be shown explicitly in \S 5.2).  Finally, although the atomic limit $\Delta_0=0$ is not perturbatively connected to the generic case of non-zero hybridization strength, we add that the present approach recovers it exactly.  Here $\Sigma_{\sigma}(\w)=0$ and simple MF alone is exact, encompassing both the $n=0$ ($\ei>0\equiv E_{\mbox{\ssz{F}}}$) and $n=2$ ($\ei+U<0$) regimes as well as the doubly degenerate local moment state arising in the singly occupied regime where $\ei<0<\ei+U$.  This of course is just a simple consequence of the fact that we work in general with broken symmetry MF propagators; but we note that recovery of the atomic limit is not at all a trivial matter if one seeks to obtain it via conventional perturbative methods, save for the accidental case of the symmetric AIM [2] where second order perturbation theory in $U$ happens to be exact in the atomic limit.

Since the diagrams for $\Sigma_{\sigma}(\w)$ have the same functional dependence on $\{\scrG_{\sigma}\}$ as considered in [27] for the symmetric AIM, most of the specific analysis of \S s 2.2, 3.2 of [27] goes through unaltered.  We thus reprise only two required results, before focusing on the central issues of stability and symmetry restoration.  First, independent of particle-hole symmetry, the $^0\Pi^{\sigma-\sigma}(\w)$ (and hence $\Pi^{\sigma-\sigma}(\w)$) satisfy $^0\Pi^{+-}(\w)=\ ^0\Pi^{-+}(-\w)$.  Only one polarization propagator need thus be considered, say $^0\Pi^{+-}(\w)$; and for later use below, $\mbox{Im}^0\Pi^{\sigma-\sigma}(\w)\geq 0 \ \forall\  \w$, and vanishes linearly in $|\w|$ as $\w\ra 0$:
\be
\frac{1}{\pi}\mbox{Im}^0\Pi^{+-}(\w)\stackrel{\w\ra 0}{\sim}|\w|D^0_{\down}(0)D^0_{\up}(0).
\label{lowwpi}
\ee 
Second, the real/imaginary parts of $^0\Pi^{+-}(\w)$ and $\Pi^{+-}(\w)$ are related by the Hilbert transform
\be
\Pi^{+-}(\w)=\int_{-\infty}^{\infty}\frac{\rmd\w_1}{\pi}\ \frac{\mbox{Im}\Pi^{+-}(\w_1)\mbox{sgn}(\w_1)}{\w_1-\w^+}.
\label{piht}
\ee
Using this, together with $\Pi^{-\sigma\sigma}(\w_1)=\Pi^{\sigma-\sigma}(-\w_1)$, equation (\ref{se}) may be expressed as 
\be
\fl \Sigma_{\sigma}(\w)=U^2\int_{-\infty}^{\infty}\frac{\rmd \w_1}{\pi}\ \mbox{Im}\Pi^{\sigma -\sigma}(\w_1)\left[\theta(\w_1)\scrG_{-\sigma}^-(\w_1+\w)+\theta(-\w_1)\scrG_{-\sigma}^+(\w_1+\w)\right]
\label{sigofw}
\ee
where
\be
\scrG_{\sigma}^{\pm}(\w)=\int_{-\infty}^{\infty}\rmd\w_1\ \frac{D^0_{\sigma}(\w_1)\theta(\pm\w_1)}{\w-\w_1\pm \im 0^+}
\label{gpmht}
\ee
denote the one-sided Hilbert transforms such that $\scrG_{\sigma}(\w)=\scrG^+_{\sigma}(\w)+\scrG^-_{\sigma}(\w)$; and $\theta(\w)$ is the unit step function.

We turn now to the important issue of stability, namely (from equation (\ref{piht})) that $\pi\mbox{Re}\Pi^{+-}(\w=0)=\int_{-\infty}^{\infty}\rmd \w_1\ \mbox{Im}\Pi^{+-}(\w_1)/|\w_1|>0$ of necessity.  For this to be satisfied, using equation (\ref{fullpi}) and that $\mbox{Im}^0\Pi^{+-}(\w=0)=0$ (equation (\ref{lowwpi})), it follows that
\be
0<U\mbox{Re}^0\Pi^{+-}(\w=0)\le 1
\ee
is required.  An explicit expression for $\mbox{Re}^0\Pi^{+-}(0)$ can however be obtained in direct parallel to that in [27] for the symmetric AIM; and is given by
\be
U\mbox{Re}^0\Pi^{+-}(\w=0)=\frac{|\mb(e_{\ii},x)|}{|\mu|}
\label{stab2}
\ee
with $|\mb(e_{\ii},x)|$ given by equation (4.2a) (and $x=\frac{1}{2}U|\mu|$).  For (any) given $U$, the contour $U\mbox{Re}^0\Pi^{+-}(\w=0;e_{\ii},x)=1$ in the $(e_{\ii},|\mu|)$ plane encloses a region of instability where $U\mbox{Re}^0\Pi^{+-}(\w=0;e_{\ii},x)>1$ and the positivity condition $\mbox{Re}\Pi^{+-}(\w=0)>0$ is violated; outside this region by contrast, $U\mbox{Re}^0\Pi^{+-}(\w=0;e_{\ii},x)\leq 1$ and stability is guaranteed.  This is illustrated in figure 2 which, for $\tilde{U}=U/\Delta_0=4\pi$ (and the wide-band AIM in practice), shows the stability border in the $(e_{\ii},|\mu|)$ plane.

\begin{figure}
\begin{center}
\psfrag{xaxis}[bc][bc]{{\large ${\bf |\mu|}$}}
\psfrag{yaxis}[bc][bc]{{\large ${\bf e_{\ii}}$}}
\epsfig{file =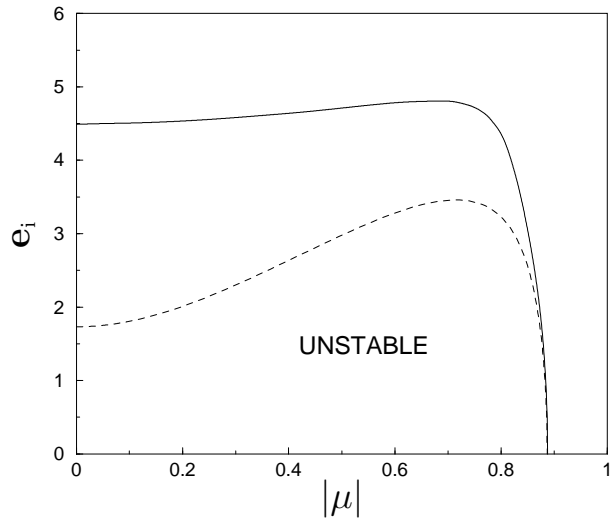,width=8cm} 
\caption{Dashed line: stability border in the $(e_{\ii},|\mu|)$ plane for fixed $\ut=4\pi$.  Solid line: corresponding solution of the symmetry restoration equation (4.3).  Full discussion in text.}
\end{center}
\end{figure}

Three important points should be noted here.  (i) First, the divergences alluded to in \S 3 that bedevil standard diagrammatic resummations based on perturbation theory in $U$ --- and thus employing restricted HF propagators (where $|\mu|=0$ is enforced) --- arise from violation of the stability condition and hence analyticity; as is evident from figure 2 for e.g.\ the symmetric AIM (corresponding to $e_{\ii}=0$ as explained above).  This deficiency, reflecting the associated instability of the restricted HF saddle point to local moment formation, does not arise in the present approach; which is why diagrammatic resummations of essentially standard form, but expressed in terms of broken symmetry MF propagators $\scrG_{\sigma}$, may be used with impunity. (ii) The stability border, illustrated in figure 2, corresponds precisely to the pure MF local moment, denoted by $|\mu_0|\equiv|\mu_0(e_{\ii})|$; for at pure MF level the local moment is obtained for any $U$ as the self-consistent solution of $|\mu_0|=|\mb(e_{\ii},\frac{1}{2}U|\mu_0|)|$ (equation (3.14a)), and hence $U^0\Pi^{+-}(\w=0;e_{\ii},x_0)=1$ from equation (4.12).  In consequence the transverse spin polarization propagator $\Pi^{+-}(\w)$ (equation (4.6)) contains a pole at $\w=0$, arising physically because the pure MF state is a degenerate doublet with no energy cost for a local spin flip.  This behaviour is correct for a local moment phase [32], and thus in the present context for the singly occupied regime in the atomic limit $\Delta_0=0$; but it is not of course correct for the ubiquitous Fermi liquid state characteristic of the metallic AIM with $\Delta_0\neq 0$, where the characteristic energy for spin flips is on the order of the Kondo scale.  The key point here however, is that this behaviour is specific {\it solely} to the pure MF level of {\it self-consistency}: from equation (4.12), $U\mbox{Re}^0\Pi^{+-}(\w=0;e_{\ii},x)=1$ {\it only} if $|\mu|$ is determined by the pure MF self-consistency equation, such that $|\mu|=|\mu_0(e_{\ii})|$.   (iii)  And within the LMA it is the symmetry restoration condition equation (4.3) that, quite generally, determines the local moment $|\mu|$ (or equivalently $x=\frac{1}{2}U|\mu|$), as explained in the previous section.  In consequence, $\mbox{Im}\Pi^{+-}(\w)$ will contain not a zero-frequency spin-flip pole but rather a resonance centred on a non-zero frequency $\w_{\m}$.  This is the Kondo scale.  Its origin within the LMA thus stems from self-consistent imposition of symmetry restoration; and since the latter guarantees Fermi liquid behaviour at low energies as explained in \S 3, its physical content is that it sets the timescale $\tau\sim h/\w_{\m}$ for restoration of the broken symmetry/degeneracy endemic at pure MF level.  The solutions of the symmetry restoration equation (4.3) in the $(e_{\ii},|\mu|)$ plane are illustrated in figure 2 for $\tilde{U}=4\pi$; and are also seen correctly to satisfy the stability criterion (as is always found in practice).  We add further that the above behaviour of $\mbox{Im}\Pi^{+-}(\w)$, arising for the AIM generically, is illustrated e.g.\ in figure 2 of [27] for the symmetric model; and that the analytic description of $\mbox{Im}\Pi^{+-}(\w)$ given by equations (2.34,35) of [27] applies {\it mutatis mutandis} to the asymmetric case.

Finally, referring to equation (4.9) for $\Sigma_{\sigma}(\w)$, it follows using $\mbox{Im}\scrG^{\pm}_{\sigma}(\w)=\mp\pi D^0_{\sigma}(\w)\theta(\pm\w)$ that
\bea
\Sigma_{\sigma}^{\subi}(\w)&=&\theta(-\w)U^2\int_0^{|\w|}\rmd \w_1\ \mbox{Im}\Pi^{\sigma -\sigma}(\w_1)D_{-\sigma}(\w_1+\w) \nonumber \\
&+&\theta(\w)U^2\int_{-|\w|}^0\rmd \w_1\ \mbox{Im}\Pi^{\sigma -\sigma}(\w_1)D_{-\sigma}(\w_1+\w).
\label{sigi}
\eea
such that $\Sigma_{\sigma}^{\subi}(\w)\geq 0$ (since $\mbox{Im}\Pi^{\sigma -\sigma}(\w)$ and $D^0_{\sigma}(\w)$ are positive semidefinite), as required by analyticity.  Using $U\mbox{Re}^0\Pi^{+-}(\w=0)<1$ (stability), it also follows from equations (4.6,7) that $\mbox{Im}\Pi^{\sigma-\sigma}(\w)$ itself vanishes linearly in $|\w|$ as $\w\ra 0$; and hence from equation (4.13) that $\Sigma^{\subi}_{\sigma}(\w)\sim\Or (\w^2)$ as $\w\ra 0$, as stated/used in \S 3 (equation (3.6b)).

\seceq
\section{Results}
We now turn to results arising from the LMA in practice, i.e.\ with $\Sigma_{\sigma}(\w)$ given approximately by equation (4.9).  The strong coupling Kondo regime, including the question of universal spectral scaling therein, is considered in \S 5.1; and the mixed valence and empty orbital regimes in \S 5.2.  The results are obtained straightforwardly: equations (4.3,4) and (3.12) are solved via the direct scheme described at the beginning of \S 4, the self-energies $\sts(\w)$ then follow via equation (4.1), and the impurity Green function and hence single-particle spectrum $D(\w)$ from equation (3.2).

\begin{figure}
\begin{center}
\psfrag{yaxis}[bc][bc]{{\large ${\bi \pi\Delta_0D(\w)}$}}
\psfrag{xaxis}[bc][bc]{{\large{\bf $\tilde{\w}$}}}
\epsfig{file =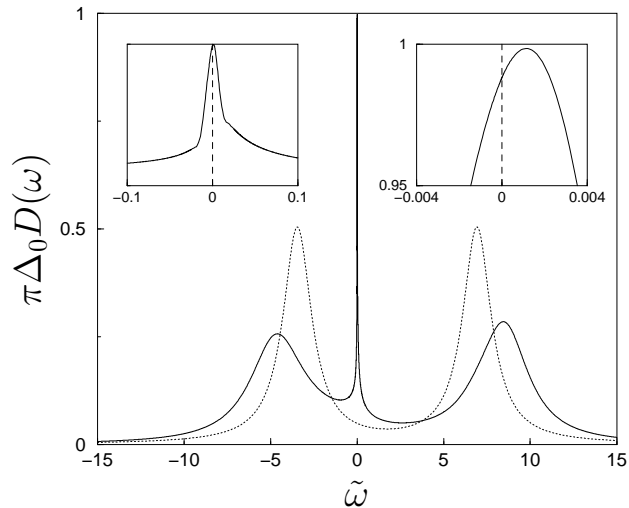,width=8cm} 
\caption{Single-particle spectrum $\pi\Delta_0D(\w)$ vs $\tilde{\w}=\w/\Delta_0$ for $\eit=\ei/\Delta_0=-4$ and $\ut=U/\Delta_0=12$ (corresponding to an asymmetry $\eta=\frac{1}{3}$).  The MF spectrum is shown for comparison (dotted line).  Insets: low-energy Kondo resonance on progressively expanded scales.}
\end{center}
\end{figure}

By way of overview figure 3 shows the resultant $\pi\Delta_0D(\w)$ vs $\tilde{\w}=\w/\Delta_0$ on all energy scales, for $\tilde{\epsilon}_{\ii}=\ei/\Delta_0=-4$ and $\ut=U/\Delta_0=12$, corresponding to an asymmetry (equation (2.2)) of $\eta=\frac{1}{3}$.  The strong coupling Kondo limit, where spectral scaling arises and the low-$\w$ behaviour of the AIM maps onto the Kondo model, corresponds strictly to $n_{\mbox{\ssz{imp}}}\ra 1$ (\S 2); although the Kondo regime is often regarded more loosely as spanning charges in the interval $0.8\lesssim n_{\mbox{\ssz{imp}}} \leq 1$ (with $\nimp=n$ for the wide-band model considered).  The example shown in figure 3 pertains to the Kondo regime, with $\nimp\simeq 0.93$; and the central Kondo resonance is shown on progressively expanded scales in the insets.  The spectral asymmetry on essentially all energy scales is evident: in the expected relative disposition of the high-energy Hubbard satellites centred on $\w\simeq -|\ei|$ and $U-|\ei|$; and in the overall shape of the Kondo resonance (left inset), including the lowest scales (right inset).  In the latter case, since the Friedel sum rule is fully satisfied, $\pi\Delta_0D(\w=0)=\sin^2[\frac{\pi}{2}\nimp]$ is correctly recovered; and since $\nimp<1$ the spectral maximum in $D(\w)$ is then pushed slightly above the Fermi level, in accordance with the quasiparticle form equation (2.7).

The qualitative deficiencies of the pure MF spectrum (also shown in figure 2) are directly apparent, even on the high-energy scales of the Hubbard satellites.  These show clearly the effects of additional many-body broadening over and above the purely elastic scattering processes captured at MF level; and the simple physical origin of which is the same as that described in [27] for the symmetric AIM, leading as seen in figure 3 to a width doubling (and peak intensity halving) of the satellites relative to pure MF.

\subsection{Kondo scaling regime}

Figure 4 illustrates the strong coupling  Kondo regime where $\nimp \ra 1$.  In figure 4a spectral evolution is shown at a fixed asymmetry $\eta= 1-2|\tilde{\epsilon}_{\ii}|/\ut=0.2$; with progressively increasing $\ut$ = 25, 30, 35 such that both Hubbard satellites move steadily outwards with a fixed ratio of their peak maxima.  In figure 4b by contrast, $\eit=-10$ is fixed and $\ut=2|\eit|/(1-\eta)$ is increased from the symmetric limit $\eta=0$ through $\eta$ = 0.1, 0.2, 0.3 and 0.4; such that the lower satellite is `frozen' while the upper, centred on $\w\sim U-|\ei|$, moves to progressively higher energies.  In both cases shown however, since $\nimp=1$ for all practical purposes, $D(\w)$ is peaked at the Fermi level $\w=0$ such that $\pi\Delta_0D(0)=1$, as evident from the insets which show the central Kondo resonance on the half-height scale.  And with progressively increasing $\ut$ in either case the Kondo scale, reflected in the width of the resonance, is rapidly decreasing.

\begin{figure}
\begin{center}
\psfrag{xaxis}[bc][bc]{{\large $\tilde{\w}$}}
\psfrag{yaxis}[bc][bc]{{\large $\pi\Delta_0D(\w)$}}
\epsfig{file =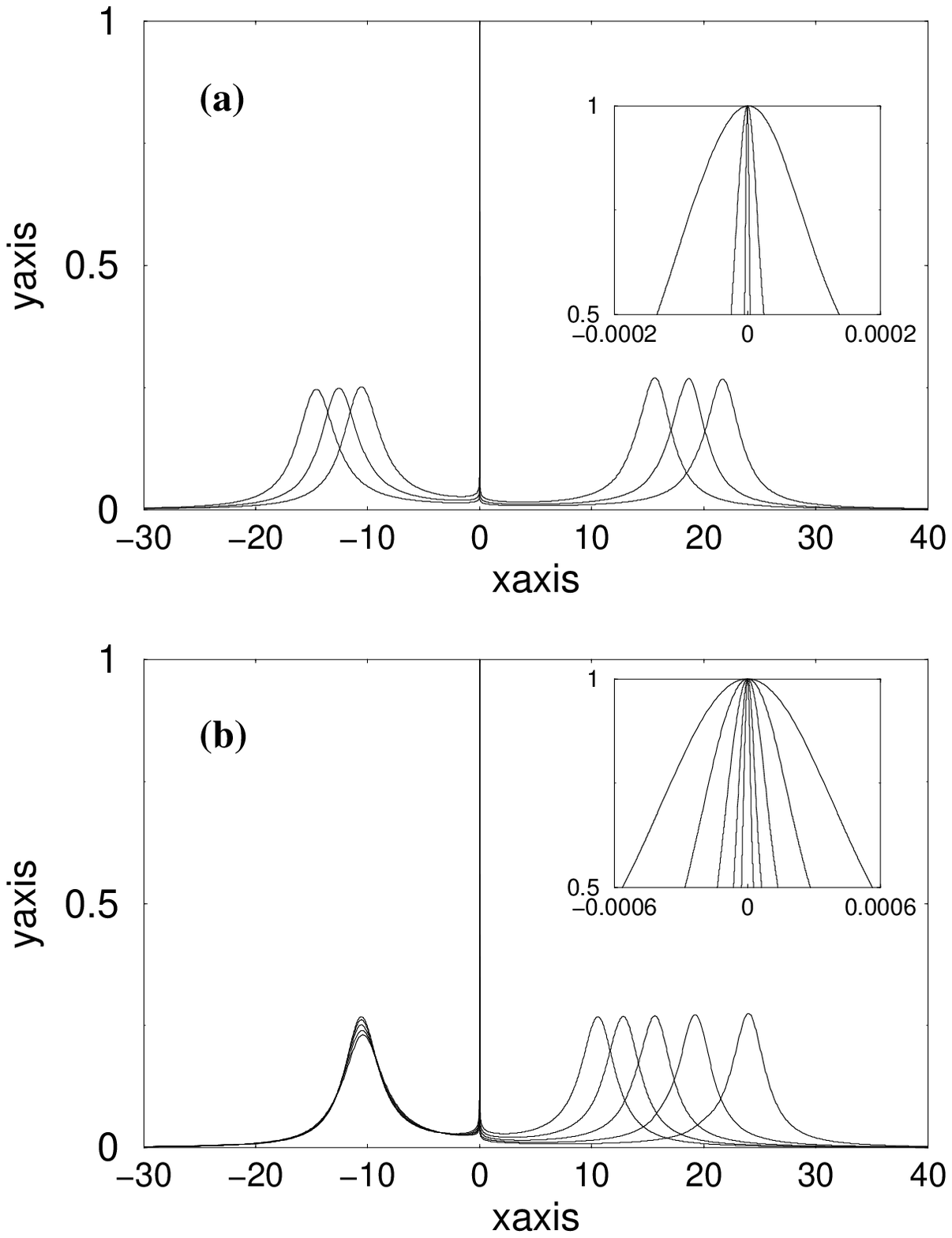,width=8cm} 
\caption{(a) $\spec$ for fixed asymmetry $\eta\equiv 1-2|\eit|/\ut=0.2$ and progressively increasing $\ut$ = 25, 30, 35.  (b) Spectral evolution for fixed $\eit=-10$ and progressively increasing $\ut=2|\eit|/(1-\eta)$ from $\eta=0$ (symmetric limit) through $\eta=$ 0.1, 0.2, 0.3, 0.4.  Insets in either case: Kondo resonances vs $\tilde{\w}$, with increasing $\ut$'s from outside to inside.}
\end{center}
\end{figure}

The first question then is the behaviour of the Kondo scale in strong coupling, equivalently the spin-flip scale $\w_{\m}$ determined (see \S 4) by symmetry restoration.  This may be obtained analytically within the present LMA; using initially, as now outlined, a generalization of the arguments detailed in [27] for the symmetric model, and with reference to equation (4.9) for $\Sigma_{\sigma}(\w)$.  In the strong coupling Kondo regime, the spectral weight of $\mbox{Im}\Pi^{+-}(\w)$ is confined entirely to $\w>0$, and $\int_0^{\infty}(\rmd \w/\pi)\ \mbox{Im}\Pi^{+-}(\w)=1$; which behaviour reflects physically the saturation of the local moment ($|\mu|\ra 1$).  The resonance in $\mbox{Im}\Pi^{+-}(\w)$ is centred on the low-energy Kondo spin-flip scale $\w_{\m}$, and on scales of this order $\scrG_{\down}^-(\w)$ is slowly varying; whence equation (4.9) for $\Sigma^{\subr}_{\up}(\w=0)$ reduces asymptotically to $\Sigma^{\subr}_{\up}(0)\sim U^2\mbox{Re}\scrG_{\down}^-(\w_{\m})$.  But $\mbox{Re}\scrG_{\down}^-(\w)$ is given by the one-sided Hilbert transform equation (4.10), which as $\w \ra 0$ is dominated by the log singularity arising necessarily because the host is metallic ($D^0_{\down}(\w=0)\neq 0$).  This $\w \ra 0$ asymptotic behaviour is captured by $\mbox{Re}\scrG_{\down}^-(\w)\sim D^0_{\down}(0)\ln [\lambda/|\w|]$; where a high-energy cutoff $\lambda$ of order min$[U,D]$ is employed ($D$ here being the host bandwidth, if one wishes to retain it as finite).  The precise value of the cutoff is immaterial in the following analysis, the key point being that the prefactor to the log divergence is precisely $D_{\down}^0(0)$.  And in strong coupling (where $x=\frac{1}{2}U|\mu| \ra \frac{1}{2}U$), $D^0_{\down}(0)$ is given asymptotically from equation (3.13) by $\pi D^0_{\down}(0)\sim \Delta_0[e_{\ii}+\frac{1}{2}U]^{-2}$.  Hence the strong coupling asymptotic behaviour of $\Sigma^{\subr}_{\up}(\w=0)$ is:
\alpheqn
\be
\Sigma^{\subr}_{\up}(\w=0)\sim\frac{\Delta_0}{\pi}\frac{U^2}{(e_{\ii}+\frac{1}{2}U)^2}\ln\left[\frac{\lambda}{\w_{\m}}\right].
\ee
A directly analogous argument may be used for $\Sigma^{\subr}_{\down}(\w=0)$, leading to $\Sigma^{\subr}_{\down}(0)\sim U^2\mbox{Re}\scrG^+_{\up}(-\w_{\m})\sim -U^2D^0_{\up}(0)\ln \left[\lambda/\w_{\m}\right]$; and hence via equation (3.13) to:
\be
\Sigma^{\subr}_{\down}(\w=0)\sim-\frac{\Delta_0}{\pi}\frac{U^2}{(e_{\ii}-\frac{1}{2}U)^2}\ln\left[\frac{\lambda}{\w_{\m}}\right].
\ee
\reseteqn
 
The $U$- and $e_{\ii}$-dependence of the Kondo scale spin-flip scale $\w_{\m}$ now follows from the symmetry restoration condition equation (4.3), which in strong coupling (where $|\bar{\mu}|\ra 1$) reduces simply to $\Sigma^{\subr}_{\up}(0)-\Sigma^{\subr}_{\down}(0)=U$; and hence using equation (5.1):
\be
\w_{\m}\sim\lambda \exp\left(\frac{-\pi}{2\Delta_0U}\frac{[e_{\ii}^2-\frac{U^2}{4}]^2}{[e_{\ii}^2+\frac{U^2}{4}]}\right).
\ee
For the symmetric AIM ($\eta=1-2|\ei|/U=0$), where $e_{\ii}=0$ (\S 4), equation (5.2) alone yields the Kondo scale; viz $\w_{\m}\sim\lambda\exp[-\pi U/8\Delta_0]$ where the exponent is exact [27].  To determine $\w_{\m}$ in general, the dependence of $e_{\ii}$ on the bare parameters $\ei$ and $U$ is naturally required.  This may be obtained using the Friedel sum rule equation (4.4) which, in the strong coupling regime where $\nimp$, $\bar{n}$ and $|\bar{\mu}|\ra 1$, gives the asymptotic behaviour $\Sigma^{\subr}_{\down}(\w=0)\sim |\ei|-U$.  Combining this with equations (5.1b,2) yields a simple quadratic for $e_{\ii}$ with solution $e_{\ii}=[U-2|\ei|]/(4f)$; where $f\equiv f(\eta^2)$ is given by
\be
f=\frac{1}{2}\left[1+(1-\eta^2)^{\frac{1}{2}}\right]
\ee
such that $f \in [1,\frac{1}{2}]$ for the $\eta \in [0,1]$ domain relevant (\S 2) to the Kondo regime.  The resultant $e_{\ii}$ is now used in equation (5.2) for the Kondo scale $\w_{\m}$, to give
\alpheqn
\bea
\w_{\m}&\sim&\lambda\exp\left(-\frac{\pi|\ei|(U-|\ei|)}{2\Delta_0U}\ \frac{1}{f}\right)\\
&=& \lambda\exp\left(-\frac{1}{2\rho J f}\right)
\eea
\reseteqn
where equation (2.13a) for $\rho J$ appropriate to the Kondo model is used.

Equation (5.4) gives the strong coupling Kondo scale for the AIM arising from the present LMA (as we have confirmed numerically).  The exponent therein differs in general, by the factor of $f(\eta^2)\in[1,\frac{1}{2}]$, from the exact result for the Kondo model; which is given to leading order in $\rho J$ by [2] $\w_{\m}\propto \exp(-1/2\rho J)$ independently of the strength of potential scattering embodied in $K/J\equiv \eta$.  As such it is exact only for the symmetric model where $\eta=0$ and $f=1$ (although we add that $f$ is slowly varying in $\eta$, lying e.g.\ within 10\% of unity for $\eta<0.6$).
Nonetheless we regard recovery of an exponentially small Kondo scale, approximate in general but close to the exact result in an obvious sense, as a non-trivial outcome of the present theory.  Moreover, provided the resultant Kondo scale is exponentially small in strong coupling (so that a clear separation between high and low energy scales arises), its precise dependence on the bare material parameters is nigh on irrelevant to the central question of scaling in the Kondo regime; viz the behaviour of the single-particle spectrum $D(\w)\equiv F(\w/\w_{\m})$ as a function of $\w/\w_{\m}$ in the formal Kondo limit $\w_{\m}\ra 0$.  This is known from previous work on the symmetric AIM [28].  Here, although the exponent in equation (5.4) for $\w_{\m}$ is exact, the prefactor $\lambda$ is simply a high-energy cutoff and certainly approximate; and yet in comparison to NRG calculations [33] the resultant universal scaling spectrum of the symmetric model is captured quantitatively by the LMA (see e.g. figures 2,3 of [28]).  It is to the questions of scaling for the asymmetric AIM that we now turn.

\begin{figure}
\begin{center}
\psfrag{yaxis}[bc][bc]{{ $\pi\Delta_0D(\w)$}}
\psfrag{xaxis}[bc][bc]{{ $\w/\w_{\m}$}}
\epsfig{file =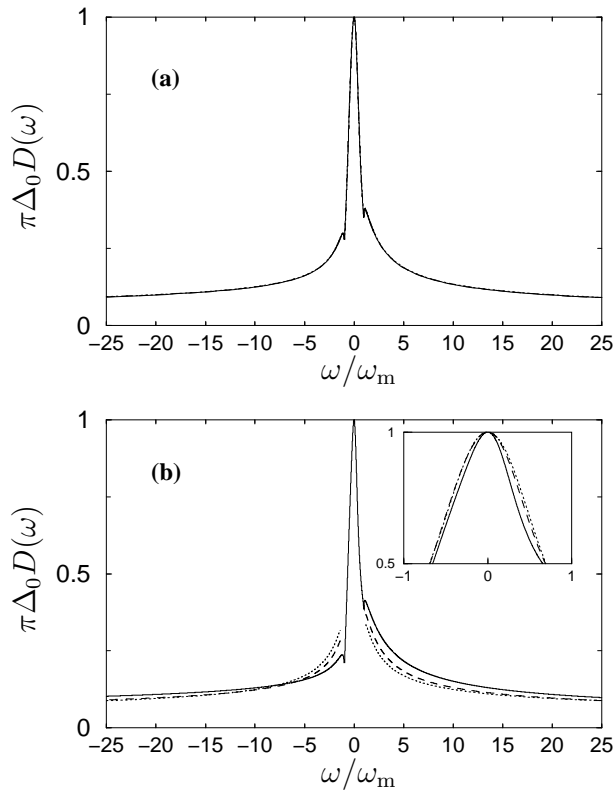,width=8cm} 
\caption{Kondo scaling regime.  (a) $\pi\Delta_0D(\w)$ vs $\w/\w_{\m}$ for fixed asymmetry $\eta\equiv 1-2|\eit|/\ut=0.2$ and increasing $\ut =$ 25, 30 , 35 (cf figure 4(a)).  The spectra collapse to a common scaling form.  (b) Resultant scaling spectrum for $\eta=0.5$ (solid line); also shown are the $|\w|/\w_{\m}\gtrsim 1$ tails of the scaling spectra for $\eta=0$ (dotted line) and $\eta=0.2$ (dashed line).  Inset: scaled Kondo resonances at higher resolution for $\eta=0$ (dotted line), 0.2 (dashed line) and 0.5 (solid line).}
\end{center}
\end{figure}

\vspace{1cm}

\noindent {\it Spectral scaling}

\vspace{0.25cm}

\noindent  The first issue in regard to spectral scaling in the Kondo regime is: does the asymmetry/potential scattering $\eta\equiv K/J$ play a role?  At first sight one might be tempted to answer no; arguing, {\it pace} for example the Kondo Hamiltonian equation (2.12), that potential scattering can be largely eliminated via a suitable canonical transformation of the host band $\{c^{\dagger}_{\kk\sigma}\}$.  And as far as the scaling spectrum is concerned this is correct at the Fermi level $\w=0$; where, as explained in \S 2 and in accordance with the quasiparticle form equation (2.7) for $G(\w)$, $\pi\Delta_0D(\w=0)=1$ provided $\nimp=1$ but regardless of how the Kondo limit $\nimp\ra 1$ is reached (i.e.\ independently of $\eta$).  Indeed if one extrapolates the quasiparticle form equation (2.7) beyond its domain of applicability as the {\it limiting} low-$\w$ behaviour of $G(\w)$, as e.g.\ in microscopic Fermi liquid theory to lowest order [2], the resultant scaling spectrum $D(\w)\equiv F(\w/\Delta_0Z)$ for $\nimp =1$ (where $\ei'=0$) is a simple Lorentzian centred on $\w =0$ and with no $\eta$-dependence; which conclusion arises also if $D(\w)$ is approximated by the local spinon spectrum [25].

If however the effects of potential scattering were entirely irrelevant, then the scaling spectrum would obviously coincide with that for the symmetric model ($\eta=0$) [33], and as such $D(\w)=D(-\w)$ would be symmetric about the Fermi level for {\it all} $\w/\w_{\m}$ (or $\w/\Delta_0Z$) and {\it any} $\eta\in [0,1]$.  Such behaviour is in our view rather implausible, and we are not aware of a convincing general argument for it.  Within the usual $U=\infty$ NCA [17-19] with $\nimp \simeq 1$ for example, the low-energy Kondo resonance is certainly asymmetric in $\w$ (albeit not correctly centred on $\w=0$ as $\nimp\ra 1$ since the Friedel sum rule is not consistently satisfied [38]).  Neither would it square readily with NRG results [39] for the asymmetric AIM, which show an evident asymmetry in the $\w$-dependence of the low-energy $D(\w)$ in the Kondo regime, see e.g.\ figure 5 of [39]; the only counterargument to which being (we believe improbably) that the latter calculations, obtained for fixed $\ut=4\pi$ by varying $\eit=-|\eit|$, are insufficiently close to the Kondo limit to eliminate asymmetry in the Kondo resonance.

If then the potential scattering/asymmetry embodied in $\eta=1-2|\eit|/\ut\equiv K/J$ is not irrelevant, the obvious question is how it influences the scaling behaviour of $D(\w)$.  Within the LMA as now shown, the answer is that a continuous family of universal scaling spectra arise; $D(\w)\equiv F(\w/\w_{\m})$ exhibiting one-parameter scaling in terms of $\w/\w_{\m}$ for each $\eta\in[0,1]$ (which is consistent with the known line of RG fixed points [9], one for each $K$).  For fixed $\eta=1-2|\eit|/\ut =0.2$, figure 4a above shows $\pi\Delta_0D(\w)$ vs the `absolute' frequency $\tilde{\w}=\w/\Delta_0$, for increasing $\ut$ = 25, 30, 35; with the corresponding behaviour of the Kondo resonances shown in the inset.  As shown in figure 5a, the Kondo resonances for the same data collapse to a common form when expressed in terms of $\w/\w_{\m}$, yielding the $\eta=0.2$ scaling spectrum  discussed further below [we add here only that, as discussed in [27,28], the small feature at $\w/\w_{\m}\simeq 1$ is entirely an artefact of the specific RPA-like form for the $\mbox{Im}\Pi^{+-}(\w)$ employed here; it can be removed, but we are content to live with it in the following].  Collapse to a scaling form occurs for each $\eta$ with increasing $\ut$, and figure 5b shows the $\eta=0.5$ scaling spectrum (solid line).  The inset thereto compares the central portion of the scaled Kondo resonance for $\eta$ = 0, 0.2 and 0.5; from which the increasing asymmetry with $\eta$ is evident, modest though it is on the scales shown.
The asymmetry induced by increasing $\eta$ is more clearly evident in the `tails' of the scaling spectra, which for $|\w|/\w_{\m} \gtrsim 1$ are also shown in figure 5b for $\eta$ = 0, 0.2 and 0.5.

The first point to note about the LMA scaling spectra is that they are not of the Lorentzian form suggested by simplistic approaches.  While the quasiparticle form is correctly recovered at sufficiently low frequencies $|\w|/\w_{\m} \ll 1$ (\S 3), the scaling spectra are by contrast dominated for $|\w|/\w_{\m}\gtrsim 1$ by the long, slowly varying tails evident in figure 5.  From a recent LMA study of the symmetric AIM in strong coupling [28], these are known to exhibit a very slow logarithmic decay, and to give excellent agreement with NRG scaling spectra [33] (see e.g.\ figure 2 of [28]).

The logarithmic tails naturally persist when $\eta\neq 0$, and in parallel to [28] it is likewise possible to determine analytically their $|\w'|=|\w|/\w_{\m}\gg 1$ asymptotic behaviour.  For $\w'=|\w'|\gg 1$ we thereby find
\be
\fl\ \ \ \ \ \pi\Delta_0D(\w)\sim\frac{1}{2}\left\{\frac{1}{[\frac{4}{\pi}g_+\ln|\w'|]^2+1}+\frac{(1+4g_-)}{[\frac{4}{\pi}g_-\ln|\w'|]^2+[1+4g_-]^2}\right\}
\ee
the (asymmetry) $\eta$-dependence of which is embodied in the factors $g_{\pm}(\eta)=f(\eta^2)[1\pm\eta]^{-1}$ (with $f$ from equation (5.3)).  For negative $\w$ and $|\w'|\gg 1$ by contrast, $\pi\Delta_0D(\w)$ is again given by the form equation (5.5), but with $g_+$ and $g_-$ simply interchanged.  In the symmetric case where $g_{\pm}(0)=1$, equation (5.5) reduces to the result compared to NRG calculations in [28].  It reproduces quantitatively the $\w'$- and $\eta$-dependences of the spectral tails shown in figure 5 for $|\w'|\gtrsim 10$ or so, and is quite satisfactory down to $|\w'|\simeq 2$; although since $\eta=1-2|\eit|/\ut\equiv K/J$ involves a ratio of `bare' parameters, the $\eta$-dependence of $g_{\pm}$ is undoubtedly approximate save for the symmetric case.

Finally, we add that since universal spectral scaling arises for fixed $\eta=1-2|\eit|/\ut$ upon increasing $\ut$, it does {\it not} therefore arise on increasing $\ut$ for fixed $|\eit|=-\eit$ (or {\it vice versa}).  For example the Kondo resonances illustrated in figure 4b for fixed $|\eit|=10$ and increasing $\ut$ do not collapse to a common scaling form when expressed as a function of $\w/\w_{\m}$, since each corresponds to a different asymmetry $\eta$.  We naturally expect the same conclusion to apply at finite $T$, namely that $D(\w;T)\equiv F(\w/\w_{\m};T/\w_{\m})$ will exhibit universal scaling only for fixed $\eta$; and hence also to the $T/\w_{\m}$ scaling behaviour of transport properties such as the resistivity, since these are determined (see e.g.\ [39]) via transport integrals in which the transport rate $\tau^{-1}(\w;T)\propto D(\w;T)$.

\subsection{Mixed valence and empty orbital regimes}
Although we have focussed latterly on scaling behaviour in the Kondo limit, the Kondo regime of the AIM is usually regarded more generally as encompassing charges in the interval $0.8\lesssim \nimp\leq 1$; while the mixed valence (MV) and empty orbital (EO) regimes of behaviour correspond respectively to $0.3\lesssim \nimp \lesssim 0.8$ and $\nimp\lesssim 0.3$.  Here we consider spectral evolution on all energy scales upon traversal of the Kondo, MV and EO regimes by progressively increasing $\eit$ from the symmetric limit $\eit=-\ut/2$; for a fixed interaction strength of $\ut=4\pi$, as considered in the NRG calculations of [39].

The Kondo regime is illustrated in figure 6 where $\pi\Delta_0D(\w)$ vs $\tilde{\w}=\w/\Delta_0$ is shown for $\eit=-\ut/2$, -4, -3 and -2 (with $\nimp\simeq 0.8$ in the latter case and the resultant charges $\nimp$ found to agree well with NRG results [39,38], to which they are compared in figure 7 below).  We regard the agreement between figure 6 from the present LMA, and the corresponding NRG results in figure 5 of [39], as rather good.  The principal differences between the two reside first in the Hubbard satellites which, we believe correctly, are more pronounced in the LMA; it being known that NRG does not do full justice to the many-body broadening of the Hubbard satellites, rendering them too diffuse [40].  And second, consistent with the expectation above, on the low-energy scale of the Kondo resonance (figure 6 inset) the LMA does not capture quantitatively the absolute value of the Kondo scale reflected e.g.\ in the resonance width; although the relative evolution of the Kondo resonances with increasing $\eit$ is respectably captured.

\begin{figure}
\begin{center}
\psfrag{yaxis}[bc][bc]{{\large $\pi\Delta_0D(\w)$}}
\psfrag{xaxis}[bc][bc]{{\large $\tilde{\w}$}}
\epsfig{file =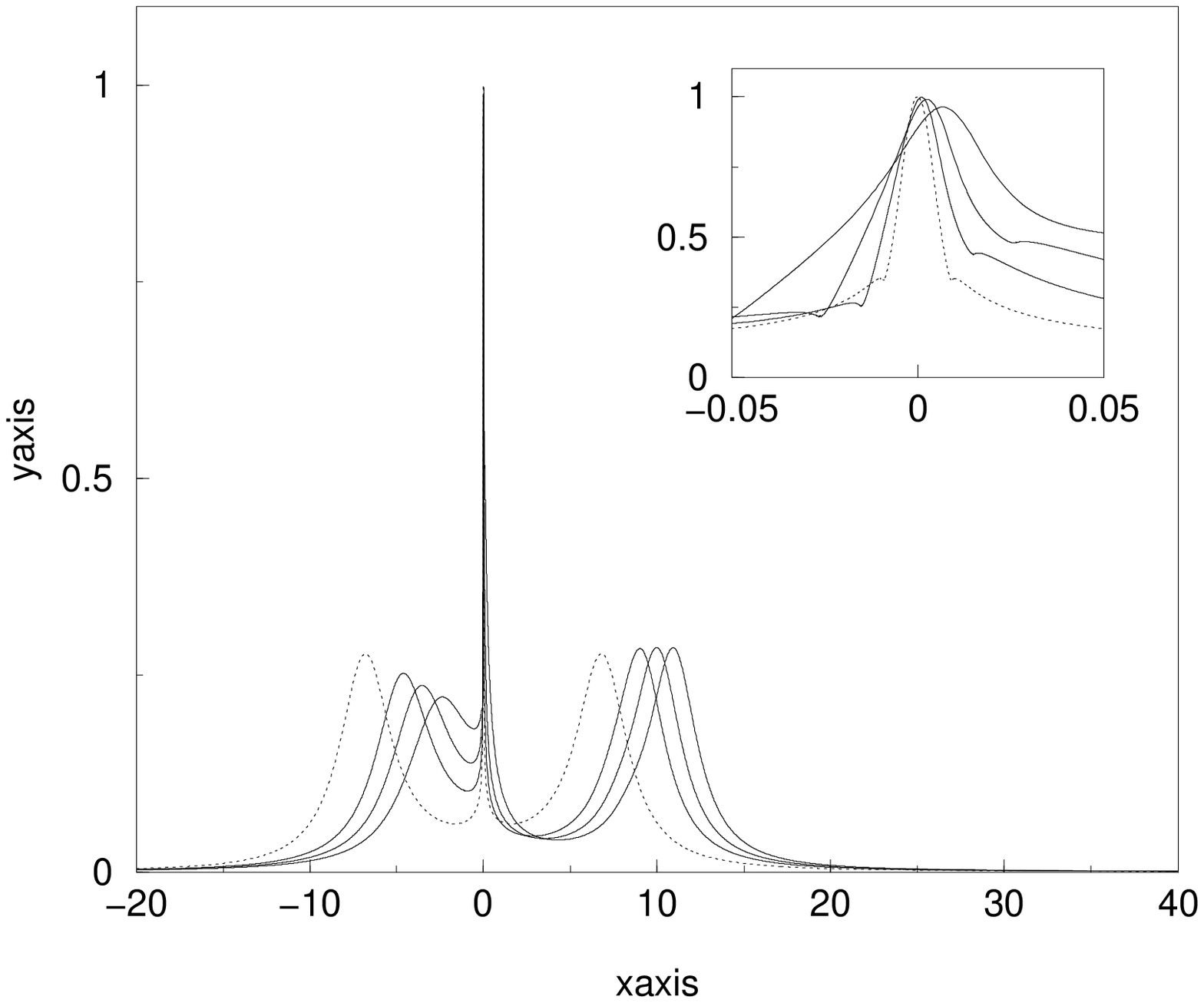,width=8cm} 
\caption{Kondo regime.  $\spec$ for fixed $\ut=4\pi$ and $\eit = -\frac{\ut}{2}$ (dotted line), -4, -3 and -2.  Inset: corresponding Kondo resonances vs $\tilde{\w}$.}
\end{center}
\end{figure}

On further increasing $\eit$ the MV regime is entered, where the impurity charge $n\equiv\nimp$ drops quite rapidly with increasing $\eit$ as the EO regime ($\eit\gtrsim 1$--2) is approached; as evident from figure 7 where the LMA $\nimp$ vs $\eit$ is shown, and compared to corresponding NRG results [38,39].  Spectral evolution in the MV and into the EO regime is illustrated in figure 8 for $\eit=-1$, 0, +1, +2, and likewise exhibits the characteristic behaviour found in NRG calculations (see e.g. figure 8 of [39]).  The Kondo regime is of course typified by a clear separation of energy scales, reflected in the exponentially small quasiparticle weight $Z\propto \w_{\m}/\Delta_0$ and hence exponentially narrow Kondo resonance.  This behaviour is lost quite rapidly on entering the MV regime, where $Z$ rises to become of order unity and the width of the Kondo resonance correspondingly becomes of order $\Delta_0$, as evident from figure 8.  This is accompanied in turn by `loss' of the lower Hubbard satellite, which becomes a barely perceptible low-energy shoulder; and by concomitant intensity erosion of the upper Hubbard satellite, although the latter remains a high energy feature centred on $\sim \ei+U$.

\begin{figure}
\begin{center}
\psfrag{yaxis}[bc][bc]{{\large $\nimp$}}
\psfrag{xaxis}[bc][bc]{{\large $\ \ \ \eit$}}
\epsfig{file =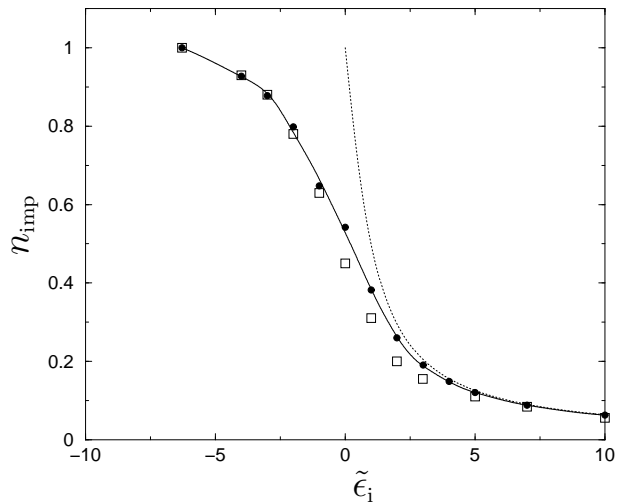,width=8cm} 
\caption{Local impurity charge $n\equiv \nimp$ vs $\eit$ for fixed $\ut =4\pi$.  Solid circles: LMA. The solid line is a guide to the eye.  Open squares: NRG results (from [39] for $\eit\leq 1$ and [38] for $\eit>1$).  The non-interacting limit charge is also shown (dotted line).}
\end{center}
\end{figure}

While the symmetry restoration condition (equation (4.3)) fundamental to the LMA is always found to be satisfied, we note that in the lower-$n$ portion of the MV regime and into the beginning of the EO regime we encounter difficulties in solving the Friedel sum rule fully self-consistently (as in the algorithm specified in \S 4).  This occurs in practice when $|\Sigma^{\subr}_{\up}(0)|$ is small ($\lesssim 0.3$); and while the reasons for it are largely technical we do not find it surprising since in this non-universal regime comparably small contributions to $\Sigma^{\subr}_{\sigma}(0)$ (in addition to the spin-fluctuation diagrams retained) will undoubtedly play a role.  In practice, as relevant to the $\eit=0$--2 examples in figure 8, we have circumvented the matter, simply by replacing $\nimp$ in equation (4.4) (or equation (3.12a)) by its MF counterpart $\bar{n}\equiv\bar{n}(e_{\ii},x)$; and although the Friedel sum rule is not then fully satisfied, the differences between $\bar{n}$ and $\nimp$ determined by spectral integration (via equation (3.12b)) are modest (e.g.\ $\sim$ 10\% for $\eit =+1$).

\begin{figure}
\begin{center}
\psfrag{yaxis}[bc][bc]{{\large $\pi\Delta_0D(\w)$}}
\psfrag{xaxis}[bc][bc]{{\large $\tilde{\w}$}}
\epsfig{file =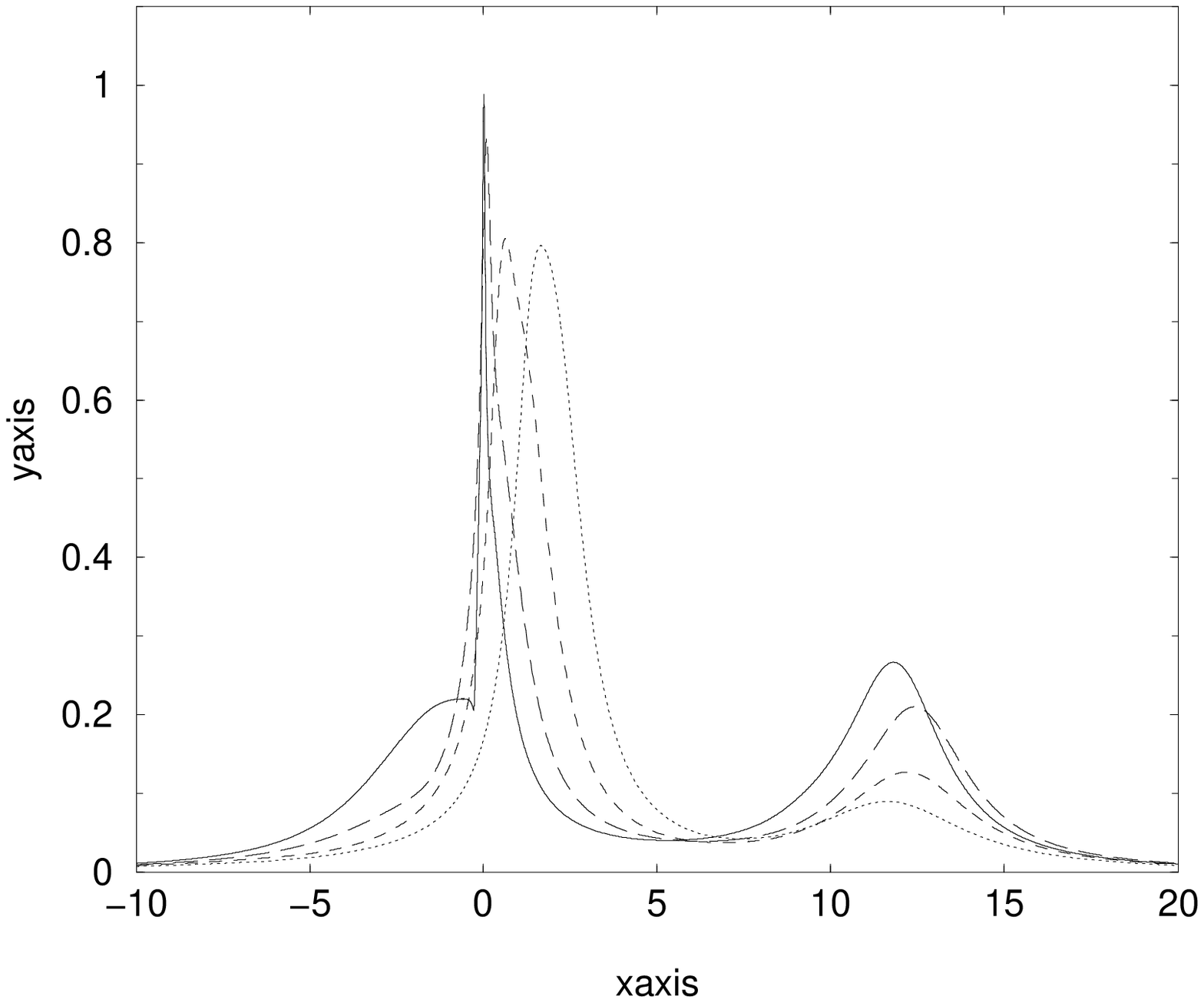,width=8cm} 
\caption{Mixed valence regime.  $\spec$ for fixed $\ut=4\pi$ and $\eit=-1$, 0, +1, +2.}
\end{center}
\end{figure}

The EO regime itself also illustrates an important facet of the LMA, for although the approach is naturally designed to capture the strong coupling physics inherent to the Kondo regime, it is nonetheless perturbatively correct in weak coupling as the non-interacting limit is approached.  This arises for fixed $\ut$ on progressively emptying the impurity site by increasing $\eit$ further in the EO regime (where we add that the Friedel sum rule is in general satisfied to full self-consistency); as evident in figure 7 where $\nimp\equiv n$ asymptotically approaches the non-interacting charge, shown as a dotted line.  Corresponding spectral evolution, $\pi\Delta_0D(\w)$ vs $\tilde{\w}=\w/\Delta_0$, is shown in figure 9 for $\eit=4$, 7 and 10.  For $\eit =4$ the upper satellite remains apparent as a weak shoulder that all but vanishes by $\eit=7$, and is entirely gone by $\eit=10$ where the resultant spectrum is seen to be near coincident with that for the non-interacting limit (dotted line).

\begin{figure}
\begin{center}
\psfrag{yaxis}[bc][bc]{{\large $\pi\Delta_0D(\w)$}}
\psfrag{xaxis}[bc][bc]{{\large $\tilde{\w}$}}
\epsfig{file =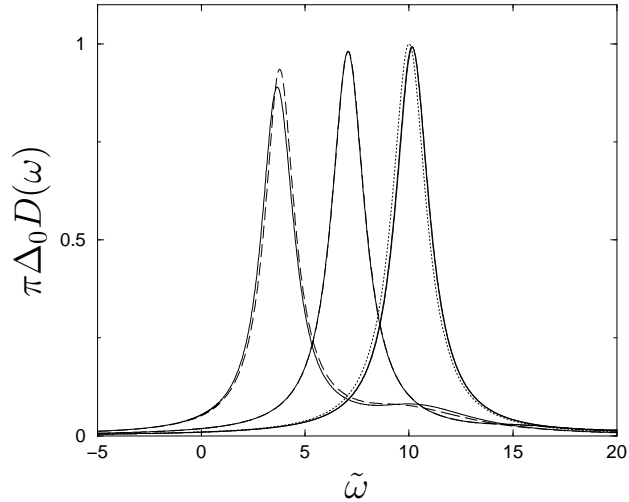,width=8cm} 
\caption{Empty orbital regime.  $\spec$ for fixed $\ut=4\pi$ and $\eit=+4$, 7, 10 (solid lines).  For $\eit=10$ the non-interacting limit spectrum is shown (dotted line).  Comparison is also made in all cases to second-order perturbation theory (SOPT, dashed line); for $\eit=7$ and 10 the LMA and SOPT spectra are indistinguishable.}
\end{center}
\end{figure}

In particular, the LMA recovers with increasing $\eit$ the weak coupling result of second-order perturbation theory (SOPT) in $U$.  The latter corresponds simply to enforcing $x=\frac{1}{2}U|\mu|=0$ (so that $\sts(\w)\equiv\Sigma(\w)$, $\scrG_{\sigma}(\w)\equiv \scrG(\w)$ etc.\ ), replacing $\Pi^{-\sigma\sigma}$ in equation (4.5) by the bare polarization bubble $^0\Pi$, and solving via steps (ii) and (iii) of the algorithm specified at the beginning of \S 4 (such that the Friedel sum rule is satisfied to full self-consistency).  For the LMA itself, $x=\frac{1}{2}U|\mu|$ is of course determined (step (i)) via symmetry restoration equation (4.3).  The resultant local moment $|\mu|$ is thereby found to diminish with increasing $\eit$  in the EO regime and vanishes for $\eit=\eito\simeq 4.3$ for the chosen $\ut=4\pi$ (far above the $\eito\simeq 0$ at which, from equation (3.15), moments vanish at pure MF level; thus ensuring, as found for the symmetric AIM [27], that physical properties evolve smoothly as $\eit$ passes through $\eito$).  For $\eit>\eito$ where $|\mu|=0$, the transverse spin polarization propagator equation (4.6) entering the LMA self-energy equation (4.5) may then be expanded perturbatively in $U$; the leading term of which is precisely the SOPT result for the self-energy.  In figure 9 LMA and SOPT single particle spectra are compared for $\eit=4$, 7, 10; the SOPT spectra being given by the dashed lines.  For $\eit=4$ the two are evidently very close (despite the persistence of moments within the LMA).  And for $\eit=7$, 10 the LMA and SOPT spectra are to all intents and purposes coincident: both are shown in the figure, but the differences are imperceptible.

\section{Summary}
We have considered in this paper single-particle dynamics of the Anderson impurity
model, a longstanding issue [1,2] pursued here via development of the local moment 
approach [27-32] to encompass the general asymmetric case. The LMA can handle
interaction strengths from strong to weak coupling, as well as dynamics on all energy
scales. And as implemented in practice, it appears to capture rather well the wide
spectrum of physical behaviour inherent to the problem; from the non-perturbative
Kondo physics characteristic of strong coupling, through mixed valence to the 
ultimately perturbative empty orbital regime. This in turn has ramifications
beyond impurity models {\it per se}, to lattice-based systems within DMFT [5-8];
for the simple, practicable approach developed here can be extended further to
encompass important problems such as e.g.\ the periodic Anderson model away from
half-filling, and hence heavy fermion physics.

\ack
We are grateful to the many people with whom we have had fruitful discussions regarding the present work.  Particular thanks are due to H R Krishnamurthy, and T Pruschke for kindly providing the NRG data used in figure 7.  We also thank the EPSRC, Leverhulme Trust and British Council for financial support.

\end{document}